\documentclass[11pt,a4paper]{article}
\pdfoutput=1 

\usepackage{jcappub}

\usepackage{amstext}
\usepackage{dcolumn}
\usepackage{bm}
\usepackage{hypcap}
\usepackage{tensor}
\usepackage{comment}
\usepackage{commath}
\usepackage[caption=false]{subfig}

\newcommand{\Planck}{\textit{Planck}}
\newcommand{\WMAP}{\textit{WMAP}}
\newcommand{\BICEP}{BICEP2/\textit{Keck}}

\newcommand{\beq}{\begin{equation}}
\newcommand{\eeq}{\end{equation}}
\renewcommand{\arraystretch}{1.6}

\def\ee{\end{equation}}
\def\bea{\begin{eqnarray}}
\def\eea{\end{eqnarray}}
\def\bse{\begin{subequations}}
\newcommand{\trm}[1]{\textnormal{#1}}
\newcommand{\Omgw}{\Omega_{\trm{gw}}}
\DeclareMathOperator{\sinc}{sinc}
\DeclareMathOperator\erf{erf}
\DeclareMathOperator\erfi{erfi}

\usepackage{xcolor}
\usepackage{soul}

\begin{document}


\title{Constraints on primordial gravitational waves from the Cosmic Microwave Background}

\author{Thomas J. Clarke,}

 \emailAdd{thomas.clarke1@nottingham.ac.uk}
\affiliation{%
 Centre for Astronomy \& Particle Cosmology, University of Nottingham, \\
 University Park, Nottingham, NG7 2RD, U.K.
}%

\author{Edmund J. Copeland,}
 \emailAdd{edmund.copeland@nottingham.ac.uk}

\author{Adam Moss}
 \emailAdd{adam.moss@nottingham.ac.uk}

\date{\today}
             
\abstract{
Searches for primordial gravitational waves have resulted in constraints in a large frequency range from a variety of sources. The standard Cosmic Microwave Background (CMB) technique is to parameterise the tensor power spectrum in terms of the tensor-to-scalar ratio, $r$, and  spectral index, $n_{\rm t}$, and constrain these using measurements of the temperature and polarization power spectra. Another method, applicable to modes  well inside the cosmological horizon at recombination, uses the shortwave approximation, under which gravitational waves behave as an effective neutrino species. In this paper we give model-independent CMB constraints on the energy density of gravitational waves, $\Omgw h^2$, for the entire range of observable frequencies. On large scales, $f \lesssim 10^{-16}\, \text{Hz}$, we reconstruct the initial tensor power spectrum in logarithmic frequency bins, finding maximal sensitivity for scales close to the horizon size at recombination. On small scales, $f \gtrsim10^{-15}\,\mbox{Hz}$, we use the shortwave approximation, finding $\Omgw h^2 < 1.7 \times10^{-6}$ for adiabatic initial conditions and $\Omgw h^2 < 2.9 \times10^{-7}$ for homogeneous initial conditions (both $2\sigma$ upper limits). For scales close to the horizon size at recombination, we use second-order perturbation theory to calculate the back-reaction from gravitational waves, finding $\Omgw h^2 < 8.4 \times10^{-7}$, in the absence of neutrino anisotropic stress and $\Omgw h^2 < 8.6 \times10^{-7}$ when including neutrino anisotropic stress. These constraints are valid for $ 10^{-15}\, \text{Hz} \gtrsim f \gtrsim 3 \times 10^{-16}\, \text{Hz}$. 
}


\maketitle
\flushbottom

\section{\label{sec:intro}Introduction}

Primordial gravitational waves (PGWs) offer a revelatory breakthrough for our knowledge of the physics of the early universe, but are currently unobserved. There are two possible sources for them: those produced during inflation, and those produced between the end of inflation and Big Bang Nucleosynthesis (BBN). For standard models of inflation, metric perturbations give rise to an almost scale invariant spectrum of PGWs, directly related to the energy scale of inflation. These result in a characteristic $B$-mode polarization signal in the Cosmic Microwave Background (CMB), which has been constrained by the \Planck\ and \BICEP\ experiments~\citep{Planck18inflation, Ade:2018gkx}. 

A number of post inflationary mechanisms could also result in the production of PGWs (see e.g.~\cite{CapriniBmodeconversion} for a recent review). Some of these processes include: (1) During the reheating phase, the non-perturbative excitation of fields can result in a stochastic background of gravitational waves, with  a well defined peak at $f \sim 10^7$-$10^8\,\text{Hz}$ (see \cite{Figueroa:2017vfa} for examples of GW production during preheating in a series of inflationary models); (2) If the curvature power spectrum has large, broad peaks on small scales, the production and merger of  Primordial Black Holes (PBHs) can lead to a  background within the range of direct detection experiments (for a recent review see \cite{Garcia-Bellido:2017fdg}); (3) A network of cosmic strings (or other topological defects) can give rise to a background at lower frequencies. There are two sources from strings: the irreducible emission from the time-evolution of the energy momentum tensor during scaling, and the production and subsequent decay of cosmic string loops. There is still some debate in the literature as to the bounds placed on the string parameters, as they vary depending how the network evolution is modelled. However, the key property that is constrained is the combination of the string tension $\mu$ and Newton's constant $G$. For the most recent bounds arising from a search in the $10^{2}$ Hz region for an isotropic stochastic background of GWs from the LIGO/VIRGO collaboration see \cite{LIGOScientific:2019vic}. Depending on the model, the bound varies between $G\mu/c^2 \leq 1.1 \times 10^{-6}$ to $G \mu/c^2 \leq 2.1 \times 10^{-14}$. A different complementary set of bounds can be obtained using the pulsar timing limits which give for the two string models, $G\mu/c^2 \leq 1.6 \times 10^{-11}$ to $G \mu/c^2 \leq 6.2 \times 10^{-12}$  \citep{29decadesofGWs}. The wide variety of processes means it is important to constrain the energy density of gravitational waves, $\Omgw h^2$, for the entire range of observable frequencies. Limits can be obtained using observations from BBN, pulsar timing, gravitational wave interferometers and the CMB. 

Measurements of the CMB temperature and polarization power spectra can be used to constrain $\Omgw h^2$.  For single-field slow-roll models of inflation, the initial spectrum of tensor perturbations is well approximated by a power-law, parameterised by the tensor-to-scalar ratio, $r$, and spectral index, $n_{\rm t}$, and is directly related to $\Omgw h^2$. This provides a limit on PGWs in the frequency range $f \lesssim 10^{-16}\, \text{Hz}$. At higher frequencies, the expected signal can be extrapolated, assuming the power-law is valid across many decades in scale. However, the assumption of a power-law across a large frequency range was shown to introduce sizeable errors on the scales probed by interferometers \cite{Giare:2020:gravitationalwavepowerlaw}. In this work we did not assume a power-law, instead directly reconstructing the tensor power spectrum in logarithmic frequency bins using the latest data from  \Planck\ and \BICEP .  

Primordial gravitational waves also have an effect on small-scale CMB anisotropies, through the back-reaction of tensor fluctuations on the cosmological background. The standard approach relies on the so-called\emph{ shortwave approximation}, which is valid for modes well inside the cosmological horizon  \cite{Taub,wheeler1962geometrodynamics,Brill_Hartle,Isaacson67,Isaacson68,misner}. Under this approximation, the energy-momentum tensor has an equation of state $w=1/3$, and so acts as an effective relativistic neutrino species.  In this paper we provide the most recent constraints using the shortwave approximation, considering both adiabatic and homogeneous initial conditions for the PGW perturbations.
 
For scales close to the horizon size at recombination, the shortwave approximation is no longer valid. No constraints on PGWs from the back-reaction of tensor fluctuations currently exist for these scales. In this work we use the approach of \cite{ABM97}, which shows that the effective energy-momentum tensor of super-Hubble modes has the form of a fluid with equation of state $w=-1/3$. There is a calculable transition period between the $w=-1/3$ super-Hubble regime and the sub-Hubble $w=1/3$ regime. Consequently a constraint can be found that isn't restricted to sub-Hubble gravitational waves.

The structure of the paper is as follows. In section~\ref{sec:previous_constraints} we review previous constraints on PGWs from CMB polarisation and from CMB temperature anisotropies. In section~\ref{sec:low_frequencies} we give model independent limits on the low-frequency reconstruction of the tensor power spectrum. Section~\ref{sec:high_frequencies} updates previous constraints that use the shortwave approximation using \Planck\ 2018 data. In section~\ref{sec:non-shortwave} a constraint is found in the intermediate regime using an expression for the gravitational wave energy momentum tensor that is valid on all scales.\footnote{We label these three regimes constrained by the CMB as low, intermediate and high frequency, although compared to other methods used to constrain gravitational waves they would all be classed as low frequency.} In section~\ref{sec:conclusions} we give conclusions.

\section{\label{sec:previous_constraints}Previous CMB constraints on PGWs}

\subsection{Constraints from B-mode polarisation}

Inflation is predicted to produce gravitational waves with both $E$ and $B$-mode polarisation, but primordial density perturbations do not result in $B$-modes. Searches for inflationary gravitational waves have therefore focused on detecting $B$-mode polarisation of the CMB (see \cite{bmodesummary} and references therein). However, there are still significant contaminants to $B$-mode observations, such as gravitational lensing along the line of sight and galactic foregrounds, which need to be accurately modelled before constraints on cosmological parameters can be found. 

When the scalar and tensor primordial power spectra have conventional power-law parameterisations, the tensor-to-scalar ratio $r_k$ is defined as the ratio of amplitudes  evaluated at scale $k$. This is used preferentially to the tensor amplitude $A_{\rm t}$ by convention, but because the scalar amplitude $A_{\rm s}$ is well determined, by for example \Planck\ \cite{Planck18}, they can be interchanged easily. The current best constraint on the tensor-to-scalar ratio is $r_{0.002}<0.056$ at $95\%$ confidence level when combining \Planck\ 2018, \BICEP\ data and Baryon Acoustic Oscillations (BAO) \citep{Planck18inflation, Ade:2018gkx}. 

A constraint on the tensor-to-scalar ratio $r$ can be converted to the gravitational wave density parameter using eq. (4) of \cite{29decadesofGWs}. In terms of $k$ and the tensor power spectrum, $\mathcal{P}_T(k)$, the gravitational wave density parameter as a function of frequency is,
\begin{equation} \label{eqn:omgw_convert}
    \Omgw (k)h^2 = \frac{3}{128} \Omega_\text{r} h^2 \mathcal{P}_T(k) \left[ \frac{1}{2} \left(\frac{k_\text{eq}}{k}\right)^2 +\frac{16}{9}\right] \,.
\end{equation}
Here $h$ contains the uncertainty in the Hubble parameter, $H_0=100 \, h\, \text{km} \,\text{s}^{-1} \text{Mpc}^{-1}$, $\Omega_\text{r}$ is the density of relativistic species and $k_\text{eq}=\sqrt{2}H_0\Omega_\text{m}/\sqrt{\Omega_\text{r}}$, for matter density $\Omega_\text{m}$, is the wavenumber of a mode that enters the horizon at matter-radiation equality. 

For single-field slow-roll models of inflation the tensor primordial power spectrum is well approximated at low-frequencies by,
\begin{equation} \label{eq:tensorpower}
    \mathcal{P}_T(k)=r A_{\rm s} \left(\frac{k}{k_*}\right)^{n_{\rm t}} \,,
\end{equation}
where the standard value for the pivot scale, $k_*=0.05 \, \text{Mpc}^{-1}$ and $n_{\rm t}$ is the tensor spectral tilt. In slow-roll models there is a consistency relation, $n_{\rm t} = -r / 8$, so inflation predicts a slightly red-tilted spectrum with $ -0.007 < n_{\rm t} < 0$ ($95\%$ confidence). 

Conversions between frequencies and wavenumbers are done using, \begin{equation}f=(1.55\times10^{-15} \, \text{Hz Mpc})\times k \,,
\end{equation}
where the numerical factor comes from the speed of light and the definition of a parsec. Using the recent \Planck\ and \BICEP\ constraint of $r_{0.002}<0.056$ , $\Omgw h^2$ ranges from $1.9 \times10^{-13}$ to $7.4\times10^{-17}$ for frequencies $3.4\times10^{-19} \, \text{Hz}$ to $2.1\times10^{-17} \, \text{Hz}$ respectively.

\subsection{Constraints from temperature anisotropies}

In the shortwave approximation, primordial gravitational waves behave like massless neutrinos and therefore contribute to the effective number of relativistic degrees of freedom, $N_\trm{eff}$. Assuming adiabatic initial conditions,  $\Omgw h^2$ can be calculated directly from $N_\trm{eff}$ as
\begin{eqnarray} \label{eq:neff}
\Omgw h^2&=& \int_0^{\infty} \dif\, (\log f) \, h^2 \, \Omgw (f) \nonumber\\ 
&\simeq& 5.6\times10^{-6}\, (N_\textnormal{eff}-3.046)=5.6\times10^{-6}\, N_\textnormal{gw} \,,
\end{eqnarray}
where it has been assumed that the standard model value $N_\trm{eff}=3.046$ holds in the absence of primordial gravitational waves. The constant in (\ref{eq:neff}) comes from the definition of $N_\textnormal{eff}$,
\begin{equation}
    \rho_\textnormal{r}=\rho_\gamma\left[1+\frac{7}{8}\left(\frac{4}{11}\right)^{4/3}N_\textnormal{eff}\right] \,,
\end{equation}
where $\rho_\textnormal{r}$ is the energy density of all relativistic species and $\rho_\gamma$ is the energy density of photons. Consequently, the conversion factor between the density parameter and the number of gravitational wave degrees of freedom is
\begin{equation}
    \frac{7}{8}\left(\frac{4}{11}\right)^{4/3} \Omega_\gamma h^2 = 5.605\times10^{-6} \,.
\end{equation}
The combination of \Planck\ 2018 + BAO data give a constraint of $N_\trm{eff} = 2.99^{+0.34}_{-0.33}$ ($2\sigma$)~\cite{Planck18}, which, using the upper limit, corresponds to $\Omgw h^2 < 1.6 \times10^{-6}$ (although a proper analysis should use $\Omgw h^2$ with a prior $\Omgw h^2 \ge 0$).

In this approximation, the perturbations of gravitational waves are identical to those of massless neutrinos, and are described by their density, $\delta_\trm{gw}$, velocity, $\theta_\trm{gw}$, shear, $\sigma_\trm{gw}$, and higher-order moments, with fluid equations given by Eqs. (\ref{eq:gwfluidequations}). It is possible that these could have non-adiabatic initial conditions, depending on the source of PGWs. Adiabatic initial conditions would be the sensible choice if primordial gravitational waves were a thermalised particle species produced by the decay of the inflaton, however most known sources of a cosmological gravitational wave background, including quantum fluctuations during inflation, reheating and cosmic strings produce an unperturbed background (see \cite{CapriniBmodeconversion} section 4.1 or \cite{maggiore_gw_book} section 22.7.2). Consequently the second choice of gravitational wave initial conditions, \textit{homogeneous} initial conditions, have no initial density perturbation (in the Newtonian gauge). In this case the gravitational wave perturbations evolve differently to the neutrino perturbations and consequently the degeneracy between $\Omgw$ and $N_\trm{eff}$ is broken. 

\cite{Smith:2006} details how CMB temperature anisotropies can be used to constrain short wavelength gravitational waves for both adiabatic and homogeneous initial conditions, using observations from \WMAP\ (first-year), \textit{SDSS} and the Lyman-$\alpha$ forest. Tighter constraints are seen for homogeneous gravitational waves compared to adiabatic gravitational waves by a factor of $\approx 5-10$. These constraints have been updated using \WMAP\ seven-year data, finding $\Omgw h^2<8.7 \times10^{-6}$ for adiabatic  and $\Omgw h^2<1.0 \times10^{-6}$ (both $2\sigma$) for homogeneous gravitational waves~\citep{ Smith_updated}. The adiabatic result has been obtained for \Planck\ 2015 data \cite{Planck_update,Pagano,Planck_update3}, finding $\Omgw h^2<1.7 \times10^{-6}$, but the homogeneous result has not been recently updated. All of these CMB constraints are smaller but are of the same order of magnitude as BBN constraints, but are valid to lower frequencies owing to the larger horizon size by the time of recombination.

\section{\label{sec:low_frequencies}Low frequencies \texorpdfstring{$\lesssim 10^{-16}\, \text{Hz}$}{: below approximately 10 femto-Hertz}}

Rather than assume a power-law spectrum~(\ref{eq:tensorpower}), we reconstruct  $\mathcal{P}_T(k)$ in logarithmic frequency bins, between a minimum and maximum wavenumber of $\log_{10} \left[ k \,  {\rm Mpc} \right] = -3.5$ and $-0.3$ respectively, with an interval of $\Delta \log_{10} \left[ k \,  {\rm Mpc} \right] = 0.2$. Outside of this range, the tensor transfer functions have very little sensitivity, so we set $\mathcal{P}_T(k)$ to the lower and upper bin values.

We use \Planck\ 2018 data \cite{Planck18} in combination with baryon-acoustic oscillation (BAO) data from the Baryon Acoustic Oscillation Survey (BOSS) DR12 \cite{BOSS}, 6dF Galaxy Survey (6dFGS) \cite{beutler:2011:6dfgsbao} and Sloan Digital Sky Survey `main galaxy sample' (SDSS-MGS) \cite{ross:2014:sdssbao}. This corresponds to the TT,TE,EE + lowE + lensing + BAO data-set used in \cite{Planck18}. The precise \textit{Planck} likelihoods used are the TT, TE and EE spectra at $l\ge 30$, the low-$\ell$ likelihood using the \textsc{Commander} component separation algorithm \cite{Planck_maps} and the low$-l$ EE likelihood from the \textsc{SimAll} algorithm in combination with \textit{Planck} 2018 lensing. 

Although \Planck\ measured the CMB polarization over
the full sky, the sensitivity for intermediate angular scales can be improved by using results from the \BICEP\ Array, with bandpowers in the range $20 < \ell < 330$. We use the most recent analysis from~\cite{Ade:2018gkx}, which includes new data from the \textit{Keck} array at 220 GHz.

For the low-frequency reconstruction, we assume an otherwise standard LCDM model, with adiabatic scalar perturbations  parameterised by a power-law spectrum with scalar amplitude $A_{\rm s}$ and spectral index $n_{\rm s}$. We assume three neutrinos species, two of these massless and a single massive neutrino with mass 0.06 eV. The other model parameters are the baryon density $\omega_{\rm b} \equiv \Omega_{\rm b} h^2$, the cold dark matter density $\omega_{\rm c} \equiv \Omega_{\rm c} h^2$, the Hubble parameter $H_0$, and the optical depth to reionization $\tau$. We assume flat priors on these parameters, and marginalise over the standard nuisance parameters in the \Planck\ and \BICEP\ likelihood codes.

\begin{figure*}
\centering
\includegraphics[width=0.7\linewidth]{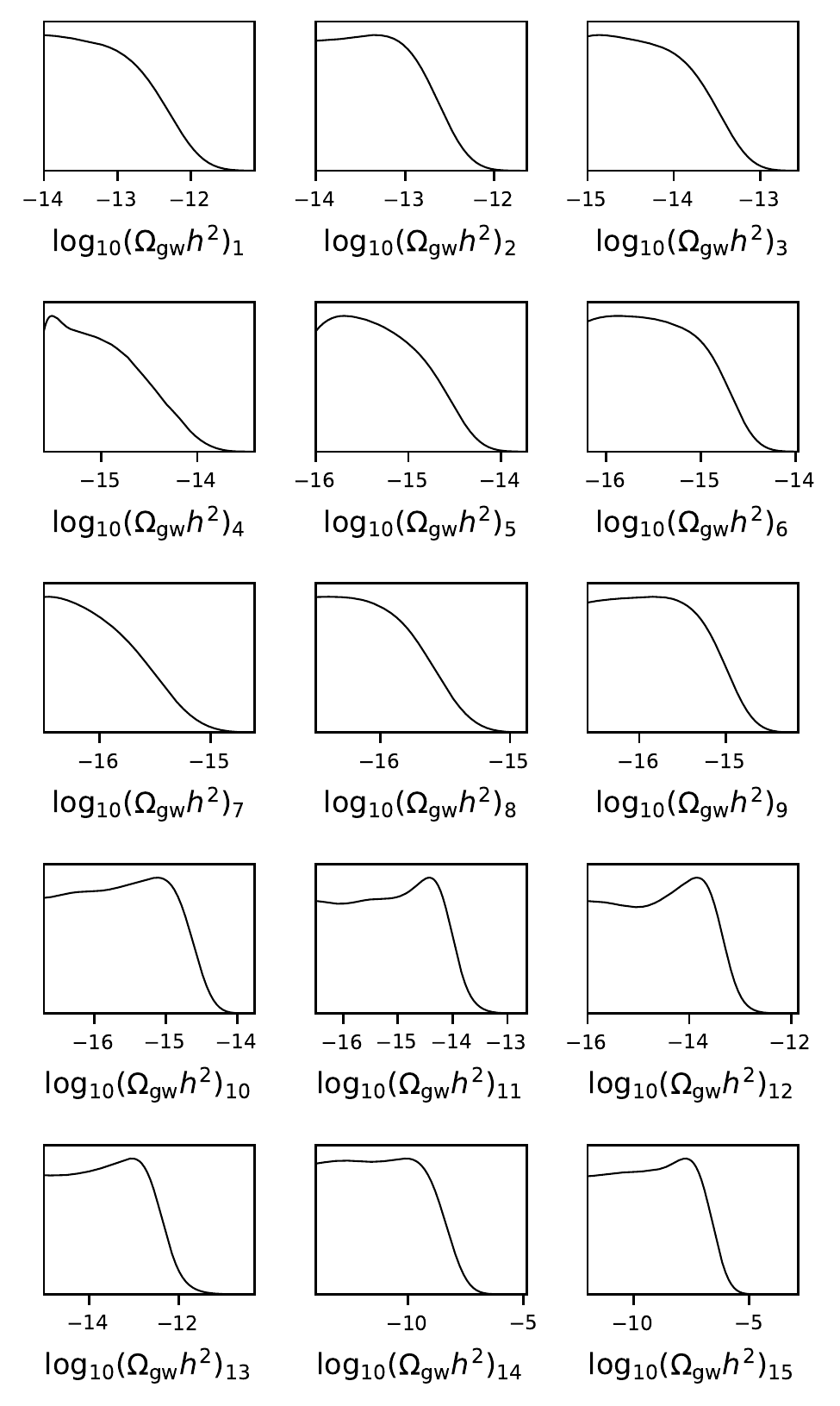}
\caption{Posterior probabilities of the gravitational wave density parameter for each logarithmic $k$-bin used for the low-frequency polarisation constraint. The bins are numbered from $\log_{10} \left[ k \,  {\rm Mpc} \right] = -3.5$ and increase in steps of $0.2$. The final bin, with $-0.5 \leq \log_{10} \left[ k \,  {\rm Mpc} \right] < -0.3$, is unconstrained and is not shown. The $95\%$ confidence limits of each posterior are used for the constraint in figure \ref{fig:density}.}
\label{fig:low_density_bins}
\end{figure*}

\begin{figure*}
\includegraphics[width=\linewidth]{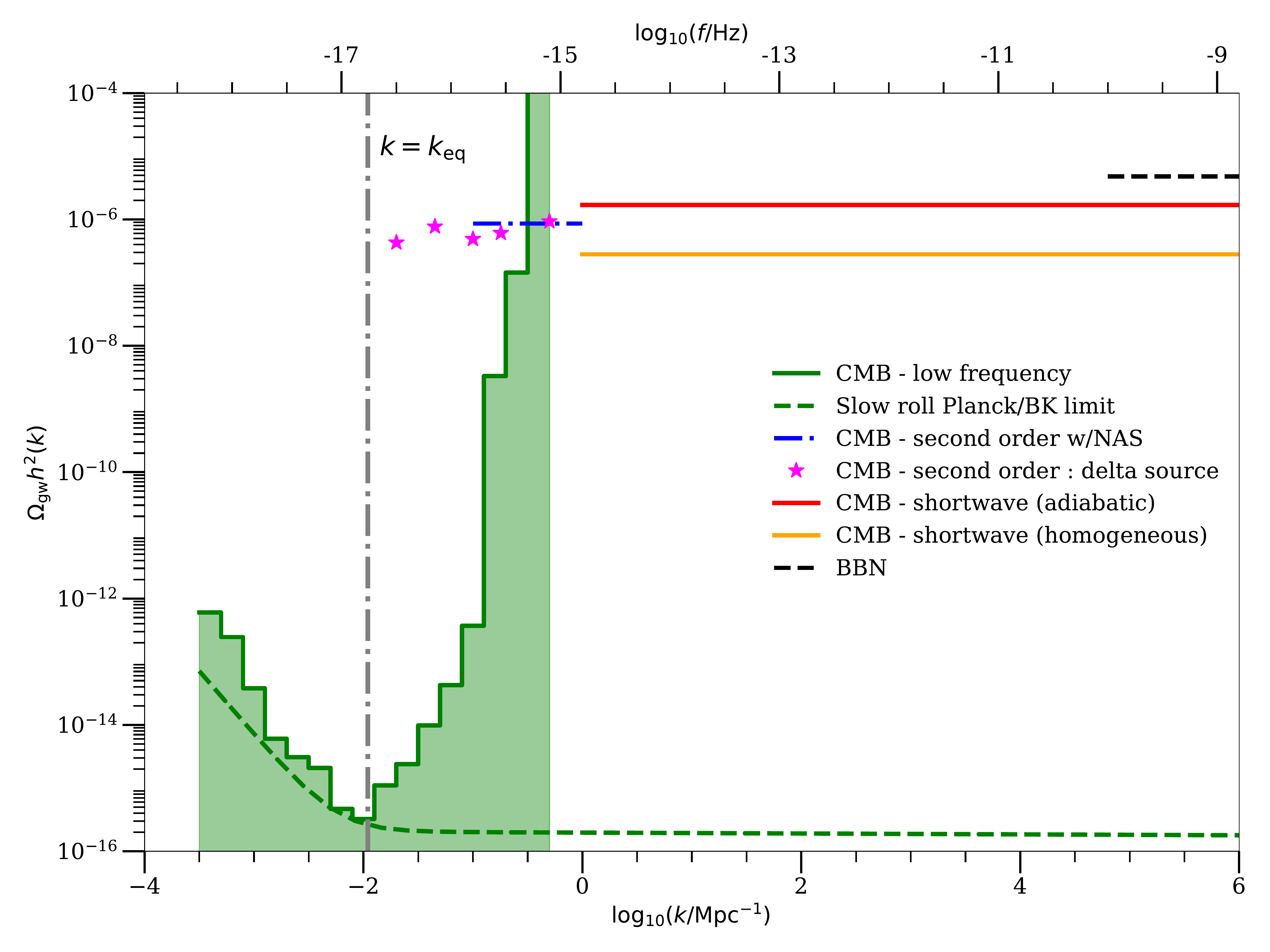}
\caption{CMB constraints on the gravitational wave density, $\Omgw h^2$, as a function of the wavenumber, $k$, and frequency, $f$. The filled green bars show constraints from a reconstruction of $\mathcal{P}_T(k)$ in logarithmic frequency bins, and the dashed green line shows a slow-roll model with the \Planck\ and \BICEP\ upper bound of $r_{0.002}=0.056$ and $n_{\rm t}=-0.007$. Shortwave constraints for adiabatic and homogeneous initial conditions are shown in red and orange (solid) respectively. The second-order back-reaction result (of section \ref{sec:low_frequencies}) for a steep source, including neutrino anisotropic stress (NAS), which we apply to scales between $k=0.1\,{\rm Mpc}^{-1}$ and $k=1\,{\rm Mpc}^{-1}$, is shown in blue (dot-dash). The constraints for delta-function sources are shown as magenta stars. A constraint from BBN \cite{BBN2} (black dashed) is also shown for comparison. The second-order back-reaction and shortwave results are integrated constraints across the given frequency range.
}
\label{fig:density}
\end{figure*}

We perform Metropolis-Hastings
Markov-chain Monte Carlo (MCMC) using a modified version of the \textsc{Cobaya} and \textsc{camb} codes~\citep{CAMB}.\footnote{\textsc{Cobaya} is available from \url{https://github.com/CobayaSampler/cobaya}.} We run four MCMC chains, stopping them when the Gelman and Rubin $R-1$ statistic is $<0.05$. The sampling is done on the power spectrum, $\mathcal{P}_T(k)$. $\Omgw(k) h^2$ is added as a derived parameter using eq. \eqref{eqn:omgw_convert} to include the variation of all the necessary variables. The posterior probabilities for each of the 16 bins are shown in figure \ref{fig:low_density_bins}. The $2\sigma$ upper limits for each of the 16 bins are used as the low-frequency constraint and are shown in figure \ref{fig:density}.
There is maximal sensitivity for scales scales close to the horizon size at recombination, and due to the decay of modes once they enter the horizon, these limits become much weaker for $f \gtrsim 10^{-16}\, \text{Hz}$. A similar result was recently found in~\cite{Namikawa:2019tax}. Above these frequencies, tighter constraints come from the second order result described in section~\ref{sec:non-shortwave}. For comparison, we also plot an inflationary model with the \Planck\ and \BICEP\ upper bound of $r_{0.002}<0.056$ and $n_{\rm t}=-0.007$.

\section{\label{sec:high_frequencies}High frequencies \texorpdfstring{$\gtrsim 10^{-15}\, \text{Hz}$}{: above approximately 1 femto-Hertz}}

The effective energy-momentum tensor for gravitational waves in the short wavelength limit is~\cite{Isaacson68,misner}
\begin{equation}
    T\indices{^{(\trm{gw})\,}_\mu^\nu}=\frac{1}{32\pi G} \langle h\indices{_{\alpha\beta|\mu}} h\indices{^{\alpha\beta|\nu}} \rangle \,,
\end{equation}
where straight lines denote covariant derivatives with respect to the background metric and $h_{\mu\nu}$ is the tensor perturbation to the conformal FLRW background metric, defined by,
\begin{equation}\label{eq:perturbed_metric}
    \dif s^2 = -a^2(\tau)\dif\tau^2+a^2(\tau)(\delta_{ij}+h_{ij})\dif x^i \dif x^j \,,    
\end{equation}
where $\tau$ is the conformal time and $a(\tau)$ is the scale factor. The angled brackets $\langle \ldots \rangle$ denote averaging over many wavelengths. We now illustrate that PGWs in this limit act like a massless neutrino species. 

The shortwave approximation (SWA) states that the perturbation is well inside the horizon and that it is oscillating much faster than the background time-scale, so we can assume that the background is approximately flat. Consequently, covariant derivatives become partial derivatives;
\begin{equation}
    T\indices{^{(\trm{gw})\,}_\mu^\nu}=\frac{1}{32\pi G} \langle h\indices{_{\alpha\beta,\mu}} h\indices{^{\alpha\beta,\nu}} \rangle \,.
\end{equation}
Furthermore, the equation of motion becomes
\begin{equation}
    h\indices{_{\mu\nu,\alpha}^\alpha}=0 \,.
\end{equation}
Considering plane waves along $z$ the solutions of the equation of motion depend on the retarded time $\tau-z$ and consequently spatial and temporal derivatives are equivalent. Therefore the trace of the energy-momentum tensor
\begin{equation}
    T\indices{^{(\trm{gw})\,}}=\frac{1}{32\pi G a^2} \left(\langle h\indices{_{ij,k}} h\indices{^{ij,k}} \rangle - \langle h\indices{_{ij,\tau}} h\indices{^{ij,\tau}} \rangle  \right) =0 \,,
\end{equation}
which implies that the equation of state $w_\trm{gw}=1/3$, as for massless neutrinos. The full solution is a sum of plane waves of positive and negative frequencies along all three spatial directions (such that the full solution is isotropic), but this argument applies to each of these spatial directions separately.

\subsection{Initial conditions}

The calculation of adiabatic initial conditions for a universe containing photons, neutrinos, cold dark matter and baryons with linear perturbations is detailed in \cite{Ma_Bertschinger} (section 7). The initial conditions for non-adiabatic modes were calculated in \cite{Bucher_Moodley_Turok}. The general methodology is to solve the perturbed Einstein and conservation equations in series solutions for small $k\tau$, where $k$ is the wavenumber of the mode under consideration and $\tau$ is the conformal time.  Matching the coefficients of the expansion gives the initial conditions for the full solution of the set of differential equations. 

This was performed in the synchronous gauge for the above system with the addition of gravitational waves with density ($\delta_\trm{gw}$), velocity ($\theta_\trm{gw}$), and shear ($\sigma_\trm{gw}$) perturbations. The gravitational wave perturbations are expanded as for the other species, for example the gravitational wave density perturbation is expanded as
\begin{equation} \label{eq:exp}
    \delta_\trm{gw}=\sum_{n=0}^\infty a^\delta_n (k\tau)^n \,,
\end{equation}
where the coefficient, $a^\delta_n$, along with the coefficients for the other species and perturbations, are the quantities we want to determine to establish the early time behaviour. We have four equations from the Einstein equations and a set of fluid conservation equations for each species \cite{Ma_Bertschinger}. For SWA gravitational waves these fluid equations are
\begin{subequations}
    \begin{align} \label{eq:gwfluidequations}
        &\dot{\delta}_\trm{gw}+\frac{4}{3}\theta_\trm{gw}+\frac{2}{3}\dot{h}=0 \,, \\
        &\dot{\theta}_\trm{gw}-\frac{1}{4}k^2(\delta_\trm{gw}-4\sigma_\trm{gw})=0 \,, \\
        &\dot{\sigma}_\trm{gw}-\frac{2}{15}(2\theta_\trm{gw}+\dot{h}+6\dot{\eta})=0 \,,
    \end{align}
\end{subequations}
where $h$ and $\eta$ are the synchronous gauge metric perturbations and dots denote differentiation with respect to conformal time.

As mentioned previously the adiabatic mode has gravitational wave perturbations identical to the neutrino perturbations. For the homogeneous mode there is one free parameter in the set of coefficients which is fixed by transforming to the Newtonian gauge using the well-known transformation relations (see \cite{Ma_Bertschinger}) and enforcing the condition that the zeroth order gravitational wave density coefficient is zero, $\tilde{a}^\delta_0=0$ (where the tilde denotes that this condition is imposed in the Newtonian gauge). The resulting adiabatic, homogeneous and gravitational wave isocurvature modes are shown in table \ref{table:initial}. The fractional contribution to the radiation density $R_i=\rho_i/\big(\sum_j \rho_j\big)$ for $i,j=\gamma,\nu,\text{gw}$. The perturbations have their standard definitions (and have a tilde in the Newtonian gauge), as do the metric perturbations. The behaviour of gravitational wave perturbations for the well-known adiabatic mode and neutrino density isocurvature mode are given along with the homogeneous gravitational wave mode and two new modes, the gravitational wave velocity isocurvature mode and the gravitational wave shear isocurvature mode.

The homogeneous mode calculated here differs from the one quoted in \cite{smiththesis} and consequently the one used in \cite{Smith:2006}.  This is for two main reasons; firstly, the gravitational wave density perturbations are not assumed to be sub-dominant to those for photons and neutrinos and secondly, and more importantly, because the homogeneous mode quoted in \cite{smiththesis} is a linear sum of the homogeneous mode given here and the neutrino density isocurvature mode. In the homogeneous mode of \cite{smiththesis} the photon and neutrino density perturbations are assumed equal. We make no such assumption but can obtain the same mode by combining the homogeneous and neutrino density isocurvature modes given in table \ref{table:initial}. The combination of these two modes can be done in the initial condition correlation matrix \cite{Bucher_Moodley_Turok} to give an equivalent result but it is the homogeneous mode given here that is the true independent mode for gravitational waves. Because of this, small differences between the results of \cite{Smith:2006} and this analysis should be expected for the homogeneous mode, even for the same data.

\begin{figure}
\capstart
\includegraphics[width=\linewidth]{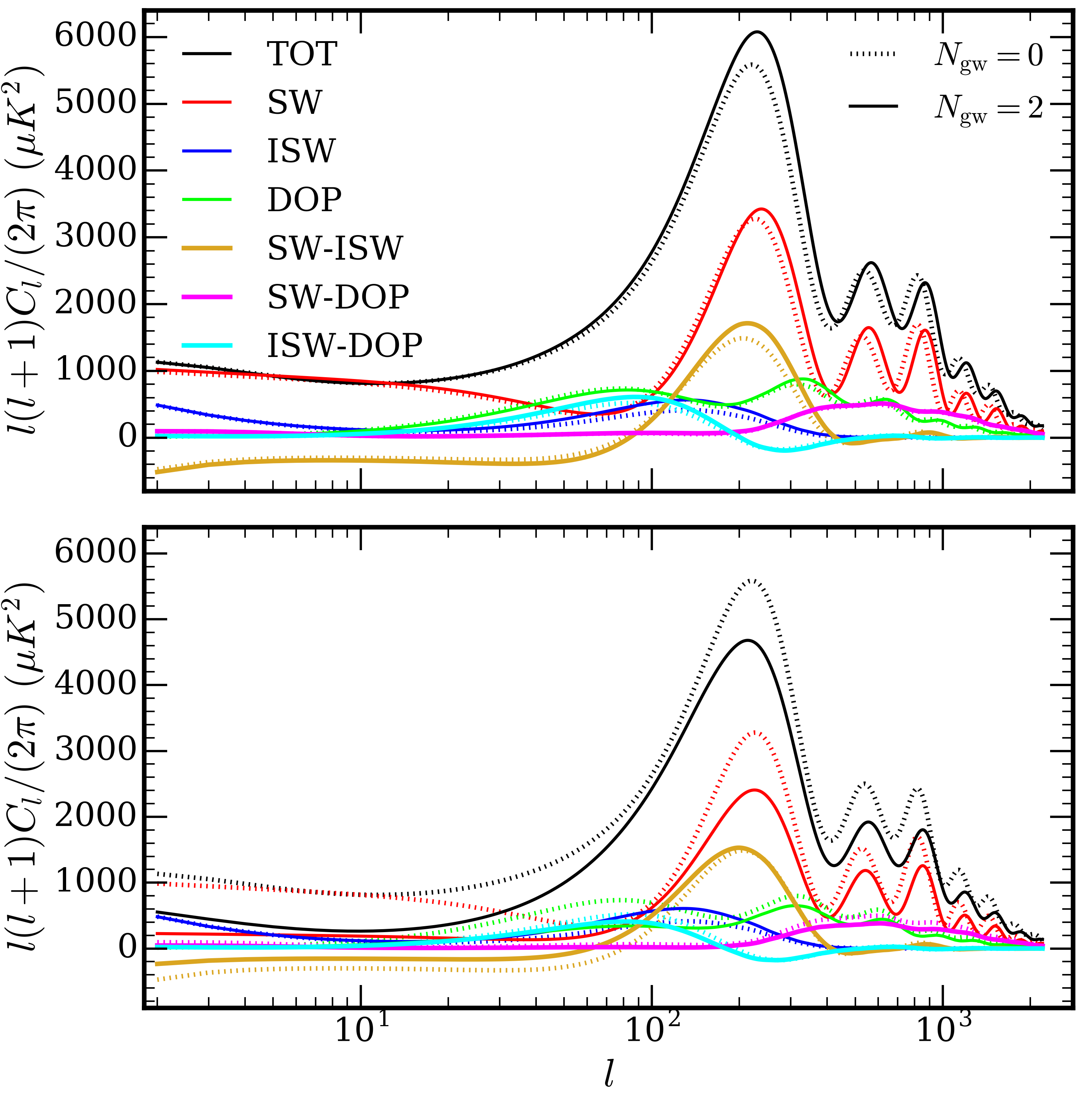}
\caption{The different contributions to the CMB power spectrum for adiabatic (top) and homogeneous (bottom) gravitational waves. The full $C_l$'s are shown along with contributions from the Sachs--Wolfe (SW), integrated Sachs--Wolfe (ISW) and Doppler terms, along with their cross-correlations. }
\label{fig:contributions}
\end{figure}

To calculate the CMB power spectrum we modified the \textsc{camb} code to include the gravitational wave equations of motion, by duplicating the massless neutrino equations, and setting the initial conditions according to table \ref{table:initial}. The changes in the CMB power spectrum when including adiabatic or homogeneous gravitational waves are shown in figure \ref{fig:contributions}. The contributions from the Sachs--Wolfe effect (SW) \cite{SachsWolfe}, integrated Sachs--Wolfe effect (ISW) and Doppler shift (DOP) are shown separately, along with the cross-correlations between them. The most noticeable difference between the adiabatic and homogeneous cases is that homogeneous gravitational waves decrease the total power whereas adiabatic gravitational waves increase the total power. This is because the homogeneous gravitational wave and photon perturbations are out of phase with each other when inside the horizon due to the initial conditions having opposite signs (see table \ref{table:initial}).

Including adiabatic gravitational waves has a very small effect on the Doppler term. Most of the change in the first peak of the power spectrum is due to the SW effect and the SW-ISW cross-correlation. This is also true for the homogeneous mode with the addition of a large change in the low-$\ell$ part of the spectrum, driven by the SW, SW--ISW and Doppler terms. The enhancement of the first peak and the decrease at low-$\ell$ is a background effect (i.e. still observable when the gravitational wave perturbations are turned off). 

It is also worth noting the behaviour of the new modes. For the gravitational wave shear isocurvature mode the neutrino and gravitational wave shear both have zero order initial conditions but they balance in such a way that the right-hand side of the relevant perturbed Einstein equation ($\propto \sum_i R_i\sigma_i$) is zero. This is also true for the density and velocity perturbations such that there are no initial metric perturbations (in synchronous or Newtonian gauge). Because of this all perturbations other than those for neutrinos and gravitational waves stay zero for all times. Using the line-of-sight integral approach of \cite{Lineofsight} it is clear that there will be no contribution to the CMB power spectrum from the shear mode.

The gravitational wave velocity isocurvature mode is the analogue of the neutrino velocity isocurvature mode and consequently behaves very similarly. In the Newtonian gauge densities and potentials have terms that go as $1/(k\tau)$. This is a consequence of the Newtonian gauge being inadequate when there is a non-zero anisotropic stress and does not mean that the perturbations diverge as $k\tau \to 0$ \cite{Bucher_Moodley_Turok}. Similarly, there is a gravitational wave density isocurvature mode not shown in table \ref{table:initial} that is a direct analogue of the neutrino density isocurvature mode. 

Baryon and cold dark matter perturbations are included in table \ref{table:initial} though the baryon velocity is not shown as $\theta_b=\theta_\gamma$, due to the tight-coupling of baryons and photons at early times \cite{Bucher_Moodley_Turok}. The baryon and dark matter perturbations behave similarly with and without gravitational waves for all modes considered here. The baryon isocurvature mode is not shown.

Isocurvature modes are well constrained by current CMB observations such that we will only consider constraints to the adiabatic and homogeneous modes here \cite{Planck18inflation}. 


\newcommand{\isoresults}[9]{ #1 & #8 & #9 & #7 & #6 & #4 \\}

\def\arraystretch{2.05} 
\setlength{\tabcolsep}{4pt} 

\begin{table*}
\capstart
\resizebox{\textwidth}{!}{\begin{tabular}{c|ccccccc}
\hline\hline

\isoresults{ }{Neut. Dens. IC}{Neut. Vel. IC}{GW. Shear IC}{Neut. Vel. IC}{GW Vel. IC}{Neut. Dens. IC} {Adiabatic} {Homogeneous}
\hline

\isoresults{$h$}{$\mathcal{O}(k^3\tau^3)$}{$\mathcal{O}(k^3\tau^3)$}{$\mathcal{O}(k^3\tau^3)$}{$\mathcal{O}(k^3\tau^3)$}{$\mathcal{O}(k^3\tau^3)$}{$\mathcal{O}(k^3\tau^3)$}{$\frac{1}{2}k^2\tau^2$}{$\frac{1}{2}k^2\tau^2$} 

\isoresults{$\eta$}{$-\frac{R_\nu}{6(19-4R_\gamma)}k^2\tau^2$}{$-\frac{4R_\nu}{3(9-4R_\gamma)}k\tau$}{$\mathcal{O}(k^3\tau^3)$}{$\mathcal{O}(k^3\tau^3)$}{$-\frac{4R_\trm{gw}}{3(9-4R_\gamma)}k\tau$}{$-\frac{R_\nu}{6(19-4R_\gamma+4R_\trm{gw})}k^2\tau^2$}{$1-\frac{(9-4R_\gamma)}{12(19-4R_\gamma)}k^2\tau^2$}{$1-\frac{(9-4R_\gamma+4R_{\trm{gw}})}{12(19-4R_\gamma+4R_{\trm{gw}})}k^2\tau^2$}

\isoresults{$\delta_\gamma$}{$-\frac{R_\nu}{R_\gamma}+\frac{R_\nu}{6R_\gamma}k^2\tau^2$}{$\frac{4R_\nu}{3R_\gamma}k\tau$}{$\mathcal{O}(k^3\tau^3)$}{$\mathcal{O}(k^3\tau^3)$}{$\frac{4R_\trm{gw}}{3R_\gamma}k\tau$}{$-\frac{R_\nu(19-4R_\gamma)}{R_\gamma(19-4R_\gamma+4R_\trm{gw})}$}{$-\frac{1}{3}k^2\tau^2$}{$-\frac{R_\trm{gw}}{R_\gamma}\frac{20}{(19-4R_\gamma+4R_\trm{gw})}$} 

\isoresults{$\theta_\gamma$}{$-\frac{R_\nu}{4R_\gamma}k^2\tau$}{$-\frac{R_\nu}{R_\gamma} k$}{$\mathcal{O}(k^4\tau^3)$}{$\mathcal{O}(k^4\tau^3)$}{$-\frac{R_\trm{gw}}{R_\gamma}k+\frac{1}{6}\frac{R_\trm{gw}}{R_\gamma}k^3\tau^2$}{$-\frac{R_\nu(19-4R_\gamma)}{4R_\gamma(19-4R_\gamma+4R_\trm{gw})}k^2\tau$}{$\mathcal{O}(k^4\tau^3)$}{$-\frac{R_\trm{gw}}{R_\gamma} \frac{5}{19-4R_\gamma+4R_{\trm{gw}}} k^2\tau$}

\isoresults{$\delta_\nu$}{$1-\frac{1}{6}k^2\tau^2$}{$-\frac{4}{3}k\tau$}{$-\frac{2R_\trm{gw}}{3R_\nu}k^2\tau^2$}{$-\frac{4}{3}k\tau$}{$\mathcal{O}(k^3\tau^3)$}{$1-\frac{1}{6}k^2\tau^2$}{$-\frac{1}{3}k^2\tau^2$}{$-\frac{1}{3}k^2\tau^2$} 

\isoresults{$\theta_\nu$}{$\frac{1}{4}k^2\tau$}{$k-\frac{13-4R_\gamma}{6(9-4R_\gamma)}k^3\tau^2$}{$\frac{R_\trm{gw}}{R_\nu}k^2\tau$}{$k-\frac{3}{10}k^3\tau^2$}{$\frac{8R_\trm{gw}}{15(9-4R_\gamma)}k^3\tau^2$}{$\frac{1}{4}k^2\tau$}{$\mathcal{O}(k^4\tau^3)$}{$\mathcal{O}(k^4\tau^3)$} 

\isoresults{$\sigma_\nu$}{$\mathcal{O}(k^3\tau^3)$}{$\frac{4}{3(9-4R_\gamma)} k\tau$}{$-\frac{R_\trm{gw}}{R_\nu}+\frac{R_\trm{gw}}{R_\nu}\frac{2}{15}k^2\tau^2$}{$\frac{4}{15}k\tau$}{$-\frac{16R_\trm{gw}}{15(9-4R_\gamma)}k\tau$}{$\frac{15+8R_\trm{gw}}{30(19-4R_\gamma+4R_\trm{gw})}k^2\tau^2$}{$\frac{2}{3(19-4R_\gamma)}k^2\tau^2$}{$\frac{2}{3(19-4R_\gamma+4R_{\trm{gw}})} k^2\tau^2$}

\isoresults{$\delta_\trm{gw}$}{-}{-}{$\frac{2}{3}k^2\tau^2$}{$\frac{4R_\nu}{3R_\trm{gw}}k\tau$}{$-\frac{4}{3}k\tau$}{$\frac{4R_\nu}{19-4R_\gamma+4R_\trm{gw}}$}{$-\frac{1}{3}k^2\tau^2$}{$\frac{20}{19-4R_\gamma+4R_{\trm{gw}}}$} 

\isoresults{$\theta_\trm{gw}$}{-}{-}{$-k^2\tau$}{$-\frac{R_\trm{gw}}{R_\nu}k+\frac{3R_\trm{gw}}{10R_\nu}k^3\tau^2$}{$k$}{$\frac{R_\nu}{19-4R_\gamma+4R_\trm{gw}}k^2\tau$}{$\mathcal{O}(k^4\tau^3)$}{$\frac{5}{19-4R_\gamma+4R_{\trm{gw}}} k^2\tau$} 

\isoresults{$\sigma_\trm{gw}$}{-}{-}{$1-\frac{2}{15} k^2\tau^2$}{$-\frac{4R_\nu}{15R_\trm{gw}}k\tau$}{$\frac{4}{15}\left(1-\frac{4R_\trm{gw}}{9-4R_\gamma}\right)k\tau$}{$\frac{R_\nu}{15(19-4R_\gamma+R_\trm{gw})}k^2\tau^2$}{$\frac{2}{3(19-4R_\gamma)}k^2\tau^2$}{$\frac{4}{3(19-4R_\gamma+4R_{\trm{gw}})} k^2\tau^2$}


\isoresults{$\delta_\trm{c}$}{-}{-}{$\mathcal{O}(k^3\tau^3)$}{-}{$\mathcal{O}(k^3\tau^3)$}{$\mathcal{O}(k^3\tau^3)$}{$-\frac{1}{4}k^2\tau$}{$-\frac{1}{4}k^2\tau$}

\isoresults{$\delta_\trm{b}$}{-}{-}{$\mathcal{O}(k^3\tau^3)$}{-}{$\frac{R_\trm{gw}}{R_\gamma}k\tau$}{$\frac{R_\nu(19-4R_\gamma)}{8R_\gamma(19-4R_\gamma+4R_\trm{gw}}k^2\tau^2$}{$-\frac{1}{4}k^2\tau$}{$-\frac{R_\gamma(19-4R_\gamma)+2R_\trm{gw}(2R_\gamma-5)}{4R_\gamma(19-4R_\gamma+4R_\trm{gw}}k^2\tau$}

\hline 

\isoresults{$\tilde{\delta}_\gamma$}{$-\frac{R_\nu(19-8R_\gamma)}{R_\gamma(19-4R_\gamma)}$}{$\frac{16R_\nu}{9-4R_\gamma} \frac{1}{k\tau}$}{$\mathcal{O}(k\tau)$}{$\mathcal{O}(k\tau)$}{$\frac{16R_\trm{gw}}{9-4R_\gamma}\frac{1}{k\tau}$}{$-\frac{R_\nu(19-8R_\gamma)}{R_\gamma(19-4R_\gamma+4R_\trm{gw})}$}{$-\frac{20}{19-4R_\gamma}$}{$-\frac{20(R_\gamma+R_\trm{gw})}{R_\gamma(19-4R_\gamma+R_\trm{gw})}$}

\isoresults{$\tilde{\theta}_\gamma$}{$-\frac{19R_\nu}{4R_\gamma(19-4R_\gamma)} k^2\tau$}{$-\frac{9R_\nu}{R_\gamma(9-4R_\gamma)} k$}{$\mathcal{O}(k^3\tau^2)$}{$\mathcal{O}(k^3\tau^2)$}{$-\frac{9R_\trm{gw}}{R_\gamma(9-4R_\nu)}k$}{$-\frac{19R_\nu}{4R_\gamma(19-4R_\gamma+4R_\trm{gw})}k^2\tau$}{$\frac{5}{19-4R_\gamma} k^2\tau$}{$\frac{5(R_\gamma-R_\trm{gw})}{R_\gamma(19-4R_\gamma+4R_\trm{gw})} k^2\tau$}

\isoresults{$\tilde{\delta}_\nu$}{$\frac{23-8R_\gamma}{19-4R_\gamma}$}{$\frac{16R_\nu}{9-4R_\gamma} \frac{1}{k\tau}$}{$\mathcal{O}(k\tau)$}{$\mathcal{O}(k\tau)$}{$\frac{16R_\trm{gw}}{9-4R_\gamma}\frac{1}{k\tau}$}{$\frac{23-8R_\gamma}{19-4R_\gamma+4R_\trm{gw}}$}{$-\frac{20}{19-4R_\gamma}$}{$-\frac{20}{19-4R_\gamma+4R_\trm{gw}}$}

\isoresults{$\tilde{\theta}_\nu$}{$\frac{15}{4(19-4R_\gamma)} k^2\tau$}{$\frac{5}{9-4R_\gamma}k$}{$\frac{R_\trm{gw}}{R_\nu}k^2\tau $}{$k $}{$-\frac{4R_\trm{gw}}{9-4R_\gamma} k$}{$\frac{15+8R_\trm{gw}}{4(19-4R_\gamma+4R_\trm{gw})} k^2\tau$}{$\frac{5}{19-4R_\gamma} k^2\tau$}{$\frac{5}{19-4R_\gamma+4R_\trm{gw}} k^2\tau$}

\isoresults{$\tilde{\delta}_\trm{gw}$}{-}{-}{$\mathcal{O}(k\tau)$}{$\mathcal{O}(k\tau)$}{$\frac{16R_\trm{gw}}{9-4R_\gamma}\frac{1}{k\tau}$}{$\mathcal{O}(k\tau)$}{$\mathcal{O}(k\tau)$}{$\mathcal{O}(k\tau)$} 

\isoresults{$\tilde{\theta}_\trm{gw}$}{-}{-}{$- k^2\tau$}{$-\frac{R_\trm{gw}}{R_\nu} k$}{$\frac{R_\trm{gw}(9-4R_\gamma-4R_\trm{gw})}{9-4R_\gamma}k$}{$-\frac{2R_\nu}{19-4R_\gamma+4R_\trm{gw}}k^2\tau$}{$\mathcal{O}(k^3\tau^2)$}{$\frac{10}{19-4R_\gamma+4R_{\trm{gw}}} k^2\tau$} 


\isoresults{$\tilde{\delta}_\trm{c}$}{-}{-}{$\mathcal{O}(k\tau)$}{-}{$\frac{16R_\trm{gw}}{(9-4R_\gamma)}\frac{1}{k\tau}$}{$\frac{4R_\nu}{19-4R_\gamma+4R_\trm{gw}}$}{$-\frac{15}{19-4R_\gamma}$}{$-\frac{20}{19-4R_\gamma+4R_\trm{gw}}$}

\isoresults{$\tilde{\theta}_\trm{c}$}{-}{-}{$\mathcal{O}(k^3\tau^2)$}{-}{-$\frac{4R_\trm{gw}}{(9-4R_\gamma)}k$}{$-\frac{R_\nu}{19-4R_\gamma+4R_\trm{gw}}k^2\tau$}{$\frac{5}{19-4R_\gamma}k^2\tau$}{$\frac{5}{19-4R_\gamma+4R_\trm{gw}}k^2\tau$}

\isoresults{$\tilde{\delta}_\trm{b}$}{-}{-}{$\mathcal{O}(k\tau)$}{-}{$\frac{16R_\trm{gw}}{(9-4R_\gamma)}\frac{1}{k\tau}$}{$\frac{4R_\nu}{19-4R_\gamma+4R_\trm{gw}}$}{$-\frac{15}{19-4R_\gamma}$}{$-\frac{20}{19-4R_\gamma+4R_\trm{gw}}$}


\isoresults{$\Phi$}{$-\frac{2R_\nu}{19-4R_\gamma}$}{$-\frac{4R_\nu}{9-4R_\gamma}\frac{1}{k\tau}$}{$\mathcal{O}(k\tau)$}{$\mathcal{O}(k\tau)$}{$\frac{4R_\gamma}{9-4R_\gamma}\frac{1}{k\tau}$}{$-\frac{2R_\nu}{19-4R_\gamma+4R_\trm{gw}}$}{$\frac{14-4R_\gamma}{19-4R_\gamma}$}{$\frac{14-4R_\gamma+4R_\trm{gw}}{19-4R_\gamma+4R_\trm{gw}}$}

\isoresults{$\Psi$}{$\frac{R_\nu}{19-4R_\gamma}$}{$\frac{4R_\nu}{9-4R_\gamma}\frac{1}{k\tau}$}{$\mathcal{O}(k\tau)$}{$\mathcal{O}(k\tau)$}{$-\frac{4R_\gamma}{9-4R_\gamma}\frac{1}{k\tau}$}{$\frac{R_\nu}{19-4R_\gamma+4R_\trm{gw}}$}{$\frac{10}{19-4R_\gamma}$}{$\frac{10}{19-4R_\gamma+4R_\trm{gw}}$}

\hline\hline
\end{tabular}}
\caption{Initial conditions on synchronous gauge (top) and Newtonian gauge quantities (bottom - with tildes) to second order in $k\tau$ (in the synchronous gauge) for the modes relevant to the gravitational wave (GW) analysis. The adiabatic and neutrino density isocurvature (IC) modes are well known and are extended here to include gravitational wave initial conditions. The homogeneous mode is the same as the homogeneous mode of \cite{smiththesis}, except for a decoupling of the neutrino density isocurvature mode, while the gravitational wave velocity IC and gravitational wave shear IC modes are new. There is a gravitational wave density IC mode that is a rescaling of the neutrino density IC mode and is not shown. The gravitational wave and neutrino shears are not given in the Newtonian gauge as the shear is gauge-invariant and hence unchanged. The baryon velocity is not shown in either gauge as $\theta_b=\theta_\gamma$, due to the tight-coupling when the initial conditions are set \cite{Bucher_Moodley_Turok} and $\theta_c$ is zero in the synchronous gauge \cite{Ma_Bertschinger}.}

\label{table:initial}
\end{table*}

\subsection{Parameter constraints}

To obtain limits on the density of gravitational waves our modified version of \textsc{camb} \cite{Lewis:1999bs} was integrated into the cosmological parameter estimation code \textsc{CosmoMC} \citep{cosmoMC} to perform an MCMC analysis. 

 We use the same data as in section~\ref{sec:low_frequencies}, but do not include tensor modes and hence \BICEP\ data. The base cosmology is an otherwise standard $\Lambda$CDM model, with the addition of $\Omgw h^2$. We obtain the following $95\%$ upper limits on the gravitational wave density parameter; 
\begin{align}
    \Omgw h^2 &< 1.7 \times10^{-6} \quad \text{(Shortwave, adiabatic)} \,, \\
    \Omgw h^2 &< 2.9 \times10^{-7} \quad \text{(Shortwave, homogeneous)} \,.
\end{align}
These constraints can be seen in figure \ref{fig:density}, along with the Big Bang Nucleosynthesis (BBN) constraint from \cite{BBN2}, the low-frequency constraint of section \ref{sec:low_frequencies} and the intermediate frequency constraint of section \ref{sec:non-shortwave}. The CMB constraints extend to much lower frequencies than those from BBN --  the exact range of validity for the shortwave approximation is discussed in section \ref{sec:shortwave-valid}. Note that these results, in contrast to the direct reconstruction, are \emph{integrated} constraints across the range of frequencies.

\section{\label{sec:non-shortwave}Intermediate frequencies  \texorpdfstring{$10^{-15}\, \text{Hz} \gtrsim  f \gtrsim 10^{-16}\, \text{Hz}$}{: between approximately 1 and 10 femto-Hertz}}

In order to consider gravitational waves without the restriction of the shortwave approximation (SWA), the work of \cite{ABM97,Brandenberger} is closely followed. Here the effective density and pressure of PGWs are calculated using the second order back-reaction of the tensor fluctuations on the metric. The second order part changes the zeroth order (or background) Einstein equations, modifying the Friedmann and continuity equations.

Expanding the Einstein equations to second order and averaging over all space (the spatial average of the linear terms are zero by definition),
\begin{equation}
    \tilde{G}\indices{^\mu_\nu} + \langle\delta^{(2)} G\indices{^\mu_\nu}\rangle = 8\pi G \left(\tilde{T}\indices{^\mu_\nu} + \langle\delta^{(2)} T\indices{^\mu_\nu}\rangle\right) \,,
\end{equation}
where $\tilde{G}\indices{^\mu_\nu}$ and $\tilde{T}\indices{^\mu_\nu}$ are the background Einstein and energy-momentum tensors respectively, and $\delta^{(2)}G\indices{^\mu_\nu}$ and $\delta^{(2)}T\indices{^\mu_\nu}$ are the second order perturbations to the Einstein and energy-momentum tensor respectively. This allows an effective energy-momentum tensor, $\tau\indices{^\mu_\nu}$ to be defined,
\begin{equation} \label{eq:emt}
    \tau\indices{^\mu_\nu} = \frac{1}{8\pi G} \left( 8\pi G \langle\delta^{(2)} T\indices{^\mu_\nu}\rangle -\langle\delta^{(2)} G\indices{^\mu_\nu}\rangle \right) \,.
\end{equation}

In general the choice of gauge is important when calculating the effective energy-momentum tensor, but since the tensor perturbation $h_{ij}$ in the transverse-traceless gauge defined in eq. (\ref{eq:perturbed_metric}) is gauge-invariant, this will not be a problem (see \cite{ABM97} for details). Consequently, in vacuum, the evaluation of the effective energy-momentum tensor for GWs simplifies to evaluating the perturbed Einstein tensor,
\begin{equation} \label{eq:emt2}
    \tau\indices{^\mu_\nu}=-\frac{1}{8\pi G} \langle\delta^{(2)}G_{\mu\nu}\rangle \,,
\end{equation}
whose components are $\tau\indices{^\mu_\nu}  = {\rm diag} \left(-\rho_\trm{gw},  \bar{p}_\trm{gw}, \bar{p}_\trm{gw}, \bar{p}_\trm{gw} \right)$. The off-diagonal terms are zero under averaging, assuming an isotropic source of PGWs. Bars are put on the pressure here as considerations of the conservation of the total energy-momentum tensor show that there is a term missing in $\bar{p}_\trm{gw}$ when calculated using eq. \eqref{eq:emt2} \cite{ABM97,Su:2012:gwenergymomentumtensoremtpressure}.

The evaluation of the effective energy-momentum tensor can be done using eq. (35.58b) of \cite{misner}. Eq. (\ref{eq:emt2}) is very similar to eq. (35.61) of \cite{misner},
\begin{equation}
    T^{(GW)}_{\mu\nu}=-\frac{1}{8\pi G}\left(\langle R^{(2)}_{\mu\nu}(h)\rangle-\frac{1}{2}\tilde{g}_{\mu\nu}\langle R^{(2)}(h)\rangle\right) \,,
\end{equation}
where $R^{(\textnormal{B})}_{\mu\nu}$ and $R^{(2)}_{\mu\nu}(h)$ are the background and second-order Ricci tensors respectively and $\tilde{g}_{\mu\nu}$ is the background metric. However, in this analysis the average is a spatial average, not over many wavelengths, so can be applied to super-horizon modes.

For the transverse-traceless perturbation defined in eq. (\ref{eq:perturbed_metric}) the components of the perturbed Einstein tensor are,
\begin{align} \label{eq:einstein_perts1}
    G^{(2)}\,\indices{^0_0}= \frac{1}{4a^2} &\left(\frac{1}{2}\dot{h}\indices{^{km}}\dot{h}\indices{_{km}} + 4\mathcal{H}h\indices{^{km}}\dot{h}\indices{_{km}} -2h\indices{^{km}}h\indices{_{km,f}^f} +h\indices{^{km,j}}h\indices{_{kj,m}}-\frac{3}{2}h\indices{^{km,j}}h\indices{_{km,j}} \right) \,, \\
    G^{(2)}\,\indices{^0_i}= \frac{1}{4a^2} &\left(-\dot{h}\indices{^{km}}h\indices{_{km,i}}- 2h\indices{^{km}}\dot{h}\indices{_{km,i}}+ 2h\indices{^{km}}\dot{h}\indices{_{ik,m}}\right) \,,  \\
    \label{eq:brandendiffer}
    G^{(2)}\,\indices{^i_0}= \frac{1}{4a^2} &\left(\dot{h}\indices{_{km}}h\indices{^{km,i}}+ 2h\indices{_{km}}\dot{h}\indices{^{km,i}}- 2h\indices{_{km}}\dot{h}\indices{^{ik,m}} +4\mathcal{H}h\indices{_{km}}h\indices{^{km,i}}-4\mathcal{H}h\indices{_{km}}h\indices{^{ik,m}} \right) \,, \\
    \label{eq:einstein_perts_4}
    G^{(2)}\,\indices{^i_j}= \frac{1}{4a^2} \Big(&-2\dot{h}\indices{^{ik}}\dot{h}\indices{_{jk}} -2h\indices{^{ik}}\ddot{h}\indices{_{jk}} +\frac{3}{2}\delta\indices{^i_j}\dot{h}\indices{^{km}}\dot{h}\indices{_{km}} +2\delta{^i_j}h\indices{^{km}}\ddot{h}\indices{_{km}} -4\mathcal{H}h\indices{^{ik}}\dot{h}\indices{_{jk}}
     \nonumber \\
    & 
    +4\mathcal{H}\delta\indices{^i_j}h\indices{^{km}}\dot{h}\indices{_{km}} -2h\indices{_{jm,k}}h\indices{^{ik,m}} +2h\indices{_{jk,m}}h\indices{^{ik,m}} -2\delta\indices{^i_j}h\indices{^{km}}h\indices{_{km,j}^j}  \nonumber \\
    & +\delta\indices{^i_j}h\indices{_{kj,m}}h\indices{^{km,j}} -\frac{3}{2}\delta\indices{^i_j}h\indices{_{km,j}}h\indices{^{km,j}} +h\indices{^{km,i}}h\indices{_{km,j}} +2h\indices{^{km}}h\indices{_{km}^{,i}_j}   \nonumber\\ & -2h\indices{^{km}}h\indices{_{jk}^{,i}_m} -2h\indices{^{km}}h\indices{^i_{k,jm}} +2h\indices{^{km}}h\indices{^i_{j,km}} +2h\indices{^{ik}}h\indices{_{jk,m}^m} \Big) \,, 
\end{align}
where we have introduced $\mathcal{H} = \dot{a}/a$, with $\dot{a} \equiv da/d\tau$.

We note that these agree with Eqs. ($8$--$11$) of \cite{Brandenberger} except for the last two terms of eq.~\ref{eq:brandendiffer}. It does however agree with eq. (2.30) of \cite{Nakamura}. This could be due to an ambiguity in the notation, to be clear, by $\dot{h}\indices{^{km,i}}$ we mean $\partial^i\partial_0 h\indices{^{km}}$. 

Because the usage of the final expressions from this section are integral to the analysis of this section of this paper the following calculation is given in detail. From the expressions for the Einstein tensor the density and pressure are found by spatial averaging defined via \cite{ABM97,Brandenberger},
\begin{equation}
    \langle A\rangle_x = \lim_{V\to\infty} \frac{1}{V} \int A \, \dif V \,,
\end{equation}
and using the gravitational wave equation of motion to find,
\begin{subequations} 
    \begin{align}
        \rho_\textnormal{gw}(\tau)=\frac{1}{8\pi Ga^2}\bigg(& \frac{1}{8}\langle\langle (\nabla h_{ij})^2\rangle\rangle_{Q,x} + \frac{1}{8}\langle\langle (\dot{h}_{ij})^2\rangle\rangle_{Q,x}  + \mathcal{H} \langle\langle h^{ij} \dot{h}_{ij}\rangle\rangle_{Q,x} \bigg) \,, \\ \label{eq:pressure_real_space}
        p_\textnormal{gw}(\tau)=\frac{1}{8\pi Ga^2}\bigg(& \frac{7}{24}\langle\langle (\nabla h_{ij})^2\rangle\rangle_{Q,x} -\frac{5}{24}\langle\langle (\dot{h}_{ij})^2\rangle\rangle_{Q,x} + \frac{\mathcal{H}}{2}(1+w^{(0)}) \langle\langle h^{ij} \dot{h}_{ij}\rangle\rangle_{Q,x} \bigg) \,, 
    \end{align}
\end{subequations}
where $w^{(0)}$ is the equation of state of the background spacetime, and we have also performed an average $Q$ over the ensemble average of stochastic initial conditions. These satisfy the continuity equation
\beq
\dot{ \rho}_\textnormal{gw} + 3 \mathcal{H} \left( \rho_\textnormal{gw} + p_\textnormal{gw} \right) = 0.    
\eeq

The background equation of state appearing in eq. \eqref{eq:pressure_real_space} is related to an important assumption. The approach of \cite{ABM97} inherently assumes that the back-reaction is a small perturbation to the background spacetime. Consequently $\rho_\text{gw} / \rho_\trm{crit}\ll1$ is required at all times. 

We next Fourier transform the tensor metric perturbation,
\begin{equation}
    h_{ij}(\vec{x},\tau)=\int \frac{\dif^3k}{(2\pi)^3}h_{ij}(\vec{k},\tau)e^{i\vec{k}.\vec{x}} \,,
\end{equation}
and decompose the Fourier components in terms of the polarisation tensor $\epsilon_{ij}$ \cite{weinberg},
\begin{equation} \label{eq:GW_helicity_expansion} 
    h_{ij}(\vec{k},\tau)=\sum_{\lambda=\pm2}\epsilon_{ij}(\hat{k},\lambda)\tilde{h}(k,\tau) \,,
\end{equation}
where $\lambda$ is the gravitational wave helicity and $\tilde{h}$ is the gravitational wave amplitude. In Fourier space the spatial averaging of a product of two general functions,
\begin{equation}
\langle f^{ij}(\vec{x},\tau)g_{ij}(\vec{x},\tau)\rangle_x=\int \frac{\dif^3k}{(2\pi)^3}\, f^{ij\,*}(\vec{k},\tau)g_{ij}(\vec{k},\tau) \,,
\end{equation}
where we have rewritten the complex exponentials from the Fourier transforms as a Dirac delta function and used this to do one of the wavenumber integrals. Here we have also set $V=1$ as it is only included to keep track of dimensions (see \cite{amendola} chapter 3). We can do this integral in spherical polar coordinates to find,
\begin{equation}
    \langle f^{ij}(\vec{x},\tau)g_{ij}(\vec{x},\tau)\rangle_x=\int \dif\, \ln{k}\,\frac{k^3}{\pi^2} \tilde{f}^*(k,\tau)\tilde{g}(k,\tau) \,,
\end{equation}
where we have evaluated the products of the polarisation tensor using \cite{weinberg},
\begin{equation} \label{eq:polel}
    \sum_\lambda \epsilon^*_{ij}(\hat{q},\lambda) \epsilon^{jk}(\hat{q},\lambda)=2\delta\indices{_i^k}-2\hat{q}_i\hat{q}^k \,.
\end{equation}
Using this for the density and pressure of gravitational waves,
\begin{subequations}
    \begin{align}
        \rho_\textnormal{gw}(\tau)&=\frac{1}{8\pi Ga^2} \int \dif\,\ln{k} \frac{k^3}{\pi^2}\tilde{\rho}_\textnormal{gw,Q}(k,\tau) \,, \\
        p_\textnormal{gw}(\tau)&=\frac{1}{8\pi Ga^2} \int \dif\,\ln{k} \frac{k^3}{\pi^2}\tilde{p}_\textnormal{gw,Q}(k,\tau) \,,
    \end{align}
\end{subequations}
where,
\begin{subequations}
    \begin{align}
        \tilde{\rho}_\textnormal{gw,Q}(k,\tau)=&\frac{k^2}{8}\langle|\tilde{h}(k,\tau)|^2\rangle_Q +\frac{1}{8}\langle|\dot{\tilde{h}}(k,\tau)|^2\rangle_Q \nonumber \\
        &+\mathcal{H}\langle|\tilde{h}^*(k,\tau)\dot{\tilde{h}}(k,\tau)|^2\rangle_Q  \,,\\
        \tilde{p}_\textnormal{gw,Q}(k,\tau)=&\frac{7k^2}{24}\langle|\tilde{h}(k,\tau)|^2\rangle_Q -\frac{5}{24}\langle|\dot{\tilde{h}}(k,\tau)|^2\rangle_Q \nonumber \\
        &+\frac{1}{2}\mathcal{H}(1+w^{(0)})\langle|\tilde{h}^*(k,\tau)\dot{\tilde{h}}(k,\tau)|^2\rangle_Q  \,.
    \end{align}
\end{subequations}

We can separate the initial condition from the time evolution of the gravitational wave amplitude as,
\begin{equation}
    \tilde{h}(k,\tau)=A_k D(k,\tau) \,,
\end{equation}
such that the primordial power spectrum,
\begin{equation}
    \mathcal{P}_\textnormal{prim}(k)=\frac{k^3}{\pi^2}\langle|A_k|^2\rangle_Q \,,
\end{equation}
specifies the ensemble average of stochastic initial conditions. 

The time evolution of the gravitational wave amplitude is now described by the function $D(k,\tau)$ which obeys the gravitational wave equation of motion,
\begin{equation} \label{eq:eom} 
    \ddot{D}+2\frac{\dot{a}}{a}\dot{D}+k^2D=16\pi Ga^2 \Pi^{\text{(T)}}  \ \,,
\end{equation}
where $\Pi^{\text{(T)}}$ is the Fourier transform of the anisotropic stress tensor decomposed in terms of the gravitational wave polarisation \cite{weinberg}. This equation comes from the first order vacuum Einstein equation for gravitational waves $R^{(1)}_{\mu\nu}(h)=0$. The helicity dependence of $\tilde{h}(k,\tau)$ was dropped earlier because this equation of motion is helicity independent.

We now have our final expressions for the density and pressure,
\begin{subequations}\label{eq:densitypressuretotal}
    \begin{align} 
                \rho_\textnormal{gw}(\tau)&=\frac{1}{8\pi Ga^2} \int_{k_{\rm min}}^{k_{\rm max}} \dif\,\ln{k}\, \tilde{\rho}_\textnormal{gw}(k,\tau) \mathcal{P}_\textnormal{prim}(k) \,, \\
        p_\textnormal{gw}(\tau)&=\frac{1}{8\pi Ga^2} \int_{k_{\rm min}}^{k_{\rm max}}  \dif\,\ln{k}\, \tilde{p}_\textnormal{gw}(k,\tau) \mathcal{P}_\textnormal{prim}(k) \,,
    \end{align}
\end{subequations}
where,
\begin{subequations} \label{eq:density_pressure} \label{eq:densitypressureD}
    \begin{align}
        \tilde{\rho}_\textnormal{gw}(k,\tau) = &\left[\frac{1}{8}\left(k^2D^2+\dot{D}^2\right) +\mathcal{H}\dot{D}D\right] \,,\\
        \tilde{p}_\textnormal{gw}(k,\tau) =  & \frac{7k^2}{24}D^2-\frac{5}{24}\dot{D}^2 +\frac{\mathcal{H}}{2}(1+w^{(0)})\dot{D}D \,,
    \end{align}
\end{subequations}
and the primordial power spectrum is conventionally parameterised via,
\begin{equation} \label{eq:powerspectrum}
    \mathcal{P}_\textnormal{prim}(k)=A_{\rm t}(k_*)\left(\frac{k}{k_*}\right)^{n_{\rm t}} \,.
\end{equation}
Consequently the methodology for calculating the density and pressure of gravitational waves is as follows. We solve the gravitational wave equation of motion (eq. (\ref{eq:eom})) for a given background cosmology and use the solution for $D(k,\tau)$ to evaluate the $k$-space density and pressure using Eqs. (\ref{eq:densitypressureD}). This is then integrated with the power spectrum of eq. (\ref{eq:powerspectrum}) to get the total homogeneous density and pressure according to Eqs. (\ref{eq:densitypressuretotal}).

\subsection{Equation of state for single fluid backgrounds} \label{sec:eos_single_fluid}
The $k$-space density and pressure can be evaluated analytically for radiation, matter and de Sitter backgrounds. This can be used to consider the behaviour of the $k$-dependent equation of state before integrating over $k$ to find the total gravitational wave density and pressure. For convenience we will define $x=k\tau$. Here the neutrino anisotropic stress is assumed to be zero.

For a radiation background, $D(x)=B \sinc{x}$, where $B$ is the initial condition on the gravitational wave amplitude, $\dot{a}/a=1/\tau$ and the equation of state of the background is $1/3$. Consequently,
\begin{align}
    \tilde{\rho}_\textnormal{gw} =  \frac{B^2}{4x^2\tau^2}\big(&-7+2x^2+7\cos{2x}+6x\sin{2x}\big) \,, \\
    \tilde{p}_\textnormal{gw} = \frac{B^2}{12x^2\tau^2}\big[&3(7-4x^2)\cos{2x} +26x\sin{2x}+2x^2-21\big] \,.
\end{align}
For a matter background, $D(x)=3B(\sinc{x}-\cos{x})/x^2$, $\dot{a}/a=2/\tau$ and the equation of state of the background is $0$. Consequently,
\begin{align}
    \tilde{\rho}_\textnormal{gw} = \frac{9B^2}{4x^6\tau^2}\big[&-39-28x^2+2x^4+(39-50x^2)\cos{2x} +6x(13-2x^2)\sin{2x}\big] \,, \\
    \tilde{p}_\textnormal{gw} = \frac{3B^2}{4x^6\tau^2}\big[&-117-56x^2+2x^4 \nonumber \\
    &+x(234-68x^2)\sin{2x} + (117-178x^2+12x^4)\cos{2x} \big] \,.
\end{align}
Finally, for a de Sitter background,
\begin{equation}
    D(x)=Bx\left(\sin{x}+\frac{\cos{x}}{x}\right)+Cx(\sinc{x}-\cos{x})\,,
\end{equation}
$\dot{a}/a=-1/\tau$ and the equation of state of the background is $-1$. Considering only the even part of $D(x)$ such that $C=0$ (though the conclusions are the same if C is included),

\begin{align}
    \tilde{\rho}_\textnormal{gw} = \frac{B^2k^2}{8}\big[&-7+2x^2 -6x\sin{2x} -7\cos{2x}\ \big] \,, \\ 
    \tilde{p}_\textnormal{gw} = \frac{B^2\tau^{-2}}{24}\big[&7+2x^2 + 14x\sin{2x}  + (7-12x^2)\cos{2x} \big]\,.
\end{align}

For the above backgrounds we can calculate the equation of state parameter $w_\trm{gw}(k,\tau)=\tilde{p}_\trm{gw}(k,\tau)/\tilde{\rho}_\trm{gw}(k,\tau)$ in the super-Hubble regime ($x\ll1$) by expanding in $x$ and in the sub-Hubble regime ($x\gg1$) by averaging trigonometric functions, e.g.,  $\langle\sin{2x}\rangle=\langle\cos{2x}\rangle=0$. Doing this we find,
\begin{equation}
    w_\trm{gw}=
    \begin{cases}
    -\frac{1}{3}, & \text{if}\ k\tau \ll 1 \,,\\
    +\frac{1}{3}, & \text{if}\ k\tau \gg 1 \,,
    \end{cases}
\end{equation}
for radiation, matter and de Sitter backgrounds.

\begin{figure}
\capstart
\includegraphics[width=\linewidth]{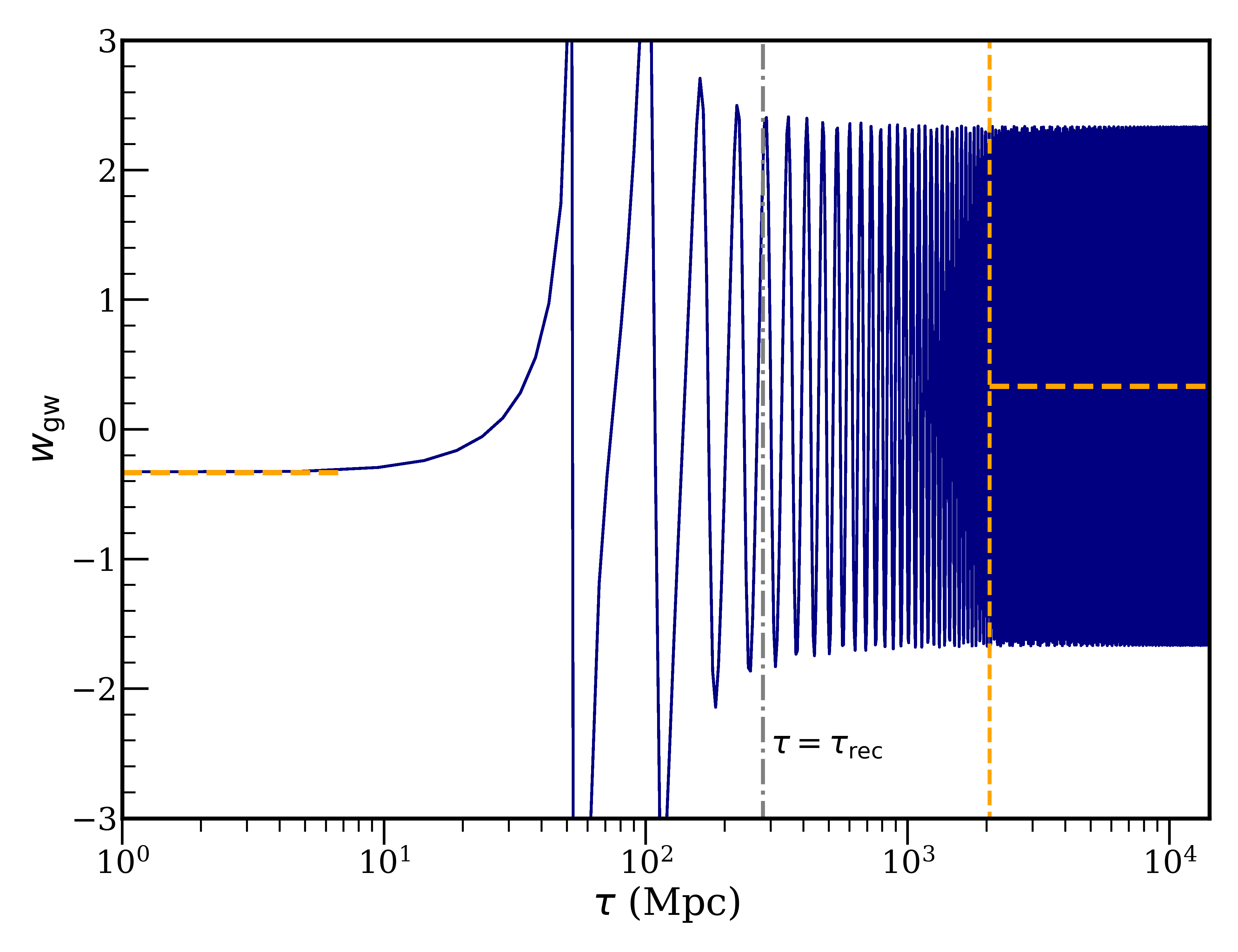}
\caption{The equation of state of gravitational waves, $w_\trm{gw}$, as a function of conformal time for a representative mode with $k=0.05 \, \text{Mpc}^{-1}$. It is $-1/3$ when the mode is outside the horizon, goes through a transition region and then oscillates about $1/3$ when well inside the horizon. The conformal time at recombination (grey dot-dash) is shown for comparison. The small and large scale values of $-1/3$ and $1/3$ are shown in orange (dashed) along with the time after which the averaging to $1/3$ is valid. }
\label{fig:eos}
\end{figure}

The equation of state for a general $\Lambda$CDM background can be solved numerically, and is shown in figure \ref{fig:eos} for \textit{Planck} 2018 parameter values and $k=0.05 \, \text{Mpc}^{-1}$. It starts at $-1/3$ when the mode is outside the horizon, then goes through a transition period where it goes through large negative and positive values before exhibiting stable oscillations about an average value of $w_\trm{gw}=1/3$.

\subsection{Shortwave validity} \label{sec:shortwave-valid}

The SWA is valid on scales averaged over a ``sufficient number of wavelengths". With a prescription how to evolve PGWs on super-horizon scales, we now quantify how many wavelengths are required to give accurate enough CMB constraints, and hence what frequency range the SWA can be applied on. 

From figure \ref{fig:eos} it is clear that there are two time-scales in the evolution of $w_\trm{gw}$: the high-frequency oscillatory behaviour, and the lower frequency transition from $-1/3$ to $1/3$. Our task is to find the  number of wavelengths required such that a constant $w_\trm{gw}=1/3$ is a good approximation to calculate the CMB power spectrum.

A cosmic variance limited CMB experiment, up to a given $\ell$, requires a precision of approximately~\citep{Seljak:2003th}
\beq
\frac{\delta C_{\ell}}{C_{\ell}} = \frac{3}{\ell}\,,
\eeq
in the power spectrum $C_{\ell}$. This corresponds to 0.1--0.2\% for $\ell=2000$. 

To proceed we assume the PGW source is a $\delta$-function for a given frequency, with energy density $\Omgw h^2$. We then compute the fractional difference in $C_{\ell}$ for the \emph{correct} evolution, compared to the SWA with constant $w_\trm{gw}=1/3$. This error decreases for higher frequency sources, as these enter the horizon and thus behave like a relativistic species earlier. We determine the minimum frequency such that $\delta C_{\ell} /C_{\ell} < 0.2\%$ for all $\ell$. Of course, it is less important to have the correct evolution  the smaller $\Omgw h^2$ is. We therefore choose $\Omgw h^2=5.6\times10^{-6}$, corresponding to $\Delta N_{\rm eff}=1$ in the SWA, such that the PGW source has an appreciable affect on the background evolution at matter-radiation equality. 

We find that a source with $k > 1\,{\rm Mpc}^{-1}$ is required for the SWA to satisfy $\delta C_{\ell} /C_{\ell} < 0.2\%$ for all $\ell$. This corresponds to the mode undergoing 50 oscillations by the epoch of equality. This limit is indicated in figure \ref{fig:eos} for a mode with a smaller $k=0.05 \, \text{Mpc}^{-1}$ --  clearly by equality a constant $w_\trm{gw}=1/3$ is not yet a good approximation. The SWA result in figure \ref{fig:density} therefore extends from $k > 1\,{\rm Mpc}^{-1}$. We note that, compared to figure 2 of \cite{Smith:2006}, they use a factor of $\sim 20$ wavelengths. Our analysis suggests that a slightly more conservative limit is required. 

\subsection{Behaviour of gravitational wave density and pressure}

\begin{figure}
\capstart
\includegraphics[width=\linewidth]{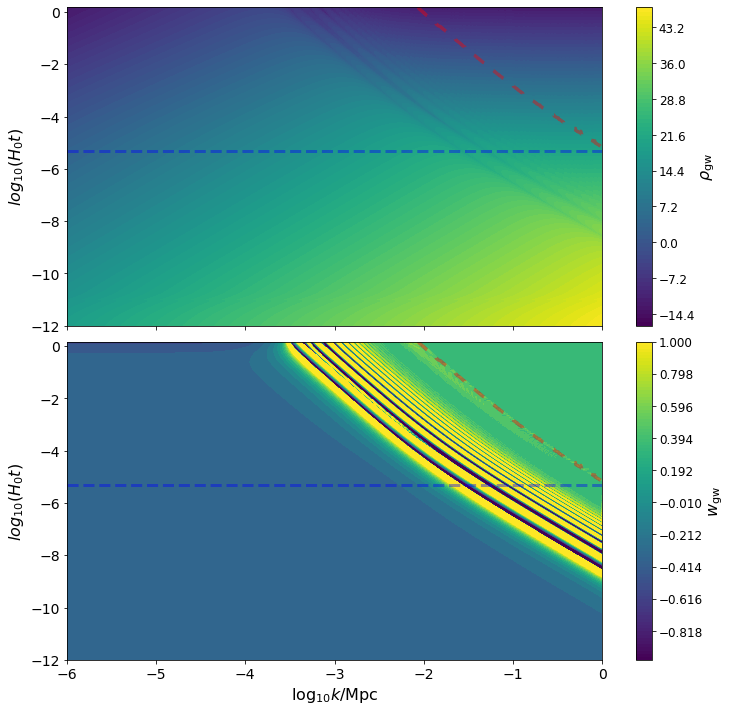}
\caption{Contour plots of the gravitational wave density, $\rho_\trm{gw}$ and equation of state, $w_\trm{gw}$ as functions of wavenumber and cosmological time for standard $\Lambda$CDM parameter values without neutrino anisotropic stress. The transition between $w_\trm{gw}=-1/3$ and $w_\trm{gw}=1/3$ can be seen clearly. The plot of the equation of state also shows an interesting feature in which super-horizon gravitational waves have an equation of state which goes below $-1/3$ during the matter to cosmological constant transition. The (red) long-dashed contour shows when each mode has undergone 50 oscillations, and the (blue) short-dashed contour the epoch of matter-radiation equality.}
\label{fig:density_w_contours}
\end{figure}

The gravitational wave density and equation of state exhibit a range of interesting physical behaviours. Figure \ref{fig:density_w_contours} shows these for standard $\Lambda$CDM parameter values as a function of $k$ and $t$ in the absence of neutrino anisotropic stress. A smoothing has been applied to $w_\trm{gw}$ to more clearly show the behaviour when the gravitational wave amplitude is highly oscillatory. The super-horizon and sub-horizon regimes can be seen clearly, along with the transition region between the two. 

When super-horizon the gravitational wave equation of state is $-1/3$ as verified above, apart from at late times, during the matter to cosmological constant transition, when it goes below $-1/3$ (this can be seen by closely inspecting the top-left of the lower panel of figure \ref{fig:density_w_contours}). This is a phenomenon that has not previously been mentioned in the literature. We analytically verified the behaviour by solving the equation of motion in a matter-cosmological constant background for small $k$ and matched solutions for different time regimes (see appendix for details). This showed that the equation of state of super-horizon modes dips below $-1/3$ during the matter to cosmological constant transition to values of $\sim -0.5$ (dependent on various parameters) but returns back to $-1/3$ soon after the cosmological constant is dominating.

\begin{figure}
\capstart
\includegraphics[width=\linewidth]{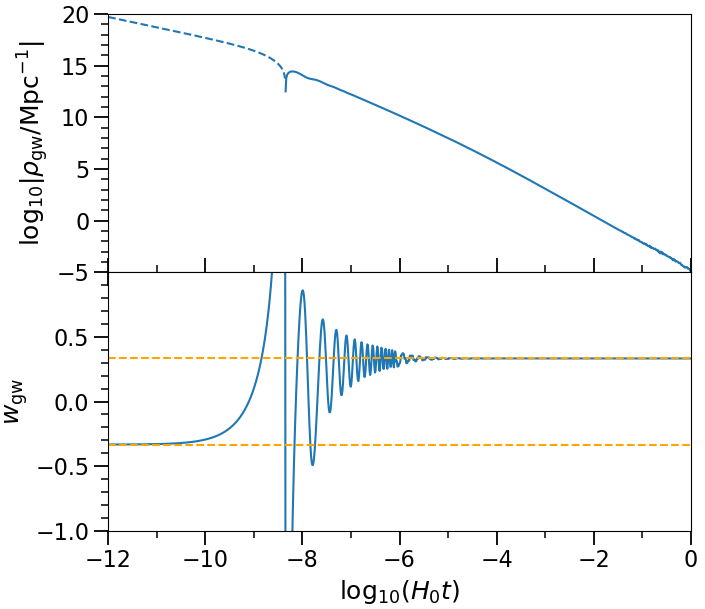}
\caption{The gravitational wave density and equation of state as a function of cosmological time after $k$-integration for a representative PGW source with $n_{\rm t}=3$, $k_{\rm min}=0.1\,{\rm Mpc}^{-1}$ and $k_{\rm max}=1\,{\rm Mpc}^{-1}$ and no neutrino anisotropic stress. The density has two regimes, one where it goes as $a^{-2}$ with a negative density (dashed) and one where it goes as $a^{-4}$ with a positive density (solid), with a transition in between. These regimes can be seen more clearly in the equation of state. There is some numerical noise as the equation of state approaches $1/3$ but this has no observable consequences.}
\label{fig:density_w_integrated}
\end{figure}

 The integrated density of eq. (\ref{eq:densitypressuretotal}) and the subsequent equation of state are shown in figure \ref{fig:density_w_integrated}, for a representative PGW source with $n_{\rm t}=3$, $k_{\rm min}=0.1\,{\rm Mpc}^{-1}$ and $k_{\rm max}=1\,{\rm Mpc}^{-1}$. The lower cutoff is chosen to be compatible with the low-frequency constraint, and the spectral index must be relatively steep, $n_{\rm t} \gtrsim 3$, to also satisfy this constraint. The high-frequency cutoff is chosen as the SWA can be used for frequencies above this. The sub- and super- Hubble regimes are clear in both cases and the transition region between the two can also be seen. 

We note that the energy density is \emph{negative} for super-Hubble modes, as stated in section 4 of \cite{Brandenberger}. Since $w_\trm{gw}=-1/3$ it can be interpreted as additional positive curvature. This contribution can lead to a reduction in the expansion rate, depending on the integration limits and spectrum in eq.~(\ref{eq:densitypressuretotal}).

\subsection{Neutrino anisotropic stress}

So far anisotropic stress has been neglected in the equation of motion for gravitational waves given in eq. \ref{eq:eom}. \cite{Weinberg:2003:gwsanisotropicstress} showed that anisotropic stress from free-streaming neutrinos has a non-negligible affect on the gravitational wave evolution.\footnote{Damping of GWs by photons has been shown to be small but can in principle be detectable via CMB spectral distortions \cite{Chluba:2014:tensorpertsphotondampingspectraldistortions}.} The neutrino anisotropic stress, $\Pi^{\text{(T)}}_\nu$ is a functional of the time derivative of the gravitational wave amplitude so the gravitational wave equation of motion becomes an integro-differential equation for the gravitational wave amplitude. The affect of this is to increase the damping term in the equation of motion and reduce the gravitational wave amplitude. This will change the analysis detailed above as, for example, the gravitational wave density and pressure are quadratic in the gravitational wave amplitude or its time derivative. 

\begin{figure}
\capstart
\includegraphics[width=\linewidth]{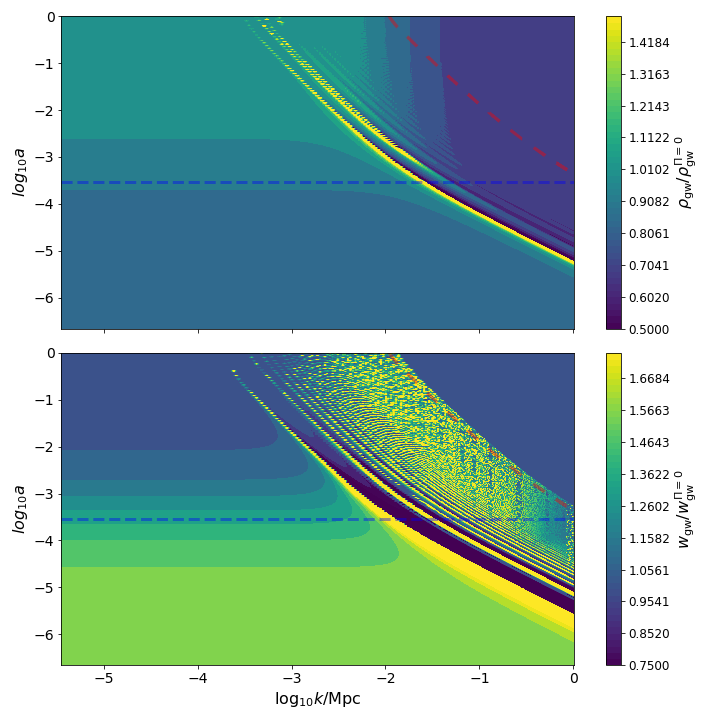}
\caption[Gravitational wave density and equation of state when including neutrino anisotropic stress]{Top panel: The ratio of the gravitational wave densities with and without neutrino anisotropic stress. Bottom panel: The ratio of the gravitational wave equation of state with and without neutrino anisotropic stress. The density roughly halves in the shortwave region but this is compensated by an equivalent reduction in the pressure such that the equation of state is still $1/3$. The equation of state becomes more negative for super-horizon scales before neutrino free-streaming. The absolute values in the absence of anisotropic stress are shown in figure \ref{fig:density_w_contours}. }
\label{fig:density_equationofstate_anisotropicstress}
\end{figure}

The change in the amplitude is most prominent when the $k$-mode comes inside the horizon. Consequently, neutrino anisotropic stress is expected to alter the behaviour of the gravitational wave density and pressure for the intermediate constraint but result in the shortwave approximation constraints still being valid.\footnote{This is neglecting the changes in the degrees of freedom in the early Universe which change the behaviour of the neutrino sector, see \cite{Watanabe:2006} for details of this which are valid in the shortwave approximation.} This is expected from the analysis of \cite{Weinberg:2003:gwsanisotropicstress}, where the sub-horizon amplitude is multiplied by a constant factor when including neutrino anisotropic stress and is confirmed in figure \ref{fig:density_equationofstate_anisotropicstress}, where the equation of state in the shortwave regime is still $1/3$ as the density and pressure both decrease by the same factor. 

The contour plots of the gravitational wave density and equation of state in figure \ref{fig:density_equationofstate_anisotropicstress} gives the ratios of these quantities in the presence and absence of anisotropic stress and shows other interesting effects. It is helpful to consider figure \ref{fig:density_w_contours} when comparing the absolute values of these quantities instead of their ratios. The equation of state of super-horizon gravitational waves before matter-radiation equality is $\approx -0.52$ and therefore considerably more negative than its value of $-1/3$ without anisotropic stress. This is due to the change in the initial condition for the time derivative of the gravitational wave amplitude when shear is included, which increases in magnitude by $\approx 1.1$.  This change in $\dot{D}$ changes the density and pressure via the third terms (which depend on $\dot{D}D$) in equations \eqref{eq:densitypressureD}.\footnote{The first terms (dependent on $k^2D^2$) are unchanged and the second terms (dependent on $\dot{D}^2$) do not contribute for super-horizon modes at early times.} The tensor initial conditions when including anisotropic stress are calculated in \cite{Shaw:2010:neutrinoinitialconditions} and show,
\begin{align}
    \dot{D}&=-\frac{5}{15+4R_\nu}k^2\tau+\mathcal{O}(k^3\tau^2) \,, \\
    \Pi^{\text{(T)}}_\nu&= \frac{4}{15+4R_\nu}k^2\tau^2+ \mathcal{O}(k^3\tau^3) \,.
\end{align}
Putting these values into Eqs. \eqref{eq:densitypressureD} and taking the ratio, the initial equation of state for super-horizon gravitational waves is,
\begin{equation}
    w_\text{gw,init}=-\frac{1}{3}\left( \frac{25+28R_\nu}{25-4R_\nu} \right ) \,.
\end{equation}
This gives $w_\text{gw,init}\approx-0.52$ for $\Lambda$CDM parameter values as seen in the numerical calculation. The equation of state increases from this value around matter-radiation equality until the super-horizon gravitational waves have an equation of state of $\approx-1/3$ after redshift $\sim 100$. The change in the equation of state for super-horizon modes during the matter-cosmological constant transition is unaffected by neutrino anisotropic stress. 

The neutrino anisotropic stress $\to0$ in the matter-dominated era which results in the density of $k$-modes being nearly unchanged by the inclusion of neutrino anisotropic stress. This can be seen above the blue-dashed line in the top panel of figure \ref{fig:density_equationofstate_anisotropicstress} and was noted in \cite{Watanabe:2006}.

\subsection{Perturbations}

The effective energy-momentum tensor at linear order is given by  
\bea \label{eqn:emtdecomp}
\tau\indices{^0_0} &=& - \left( \rho + \delta\rho \right)\,, \\
\tau\indices{^0_i} &=& \left( \rho + p \right) v_{i}\,, \\
\tau\indices{^i_j} &=& \left( p + \delta{p} \right) \delta\indices{^i_j} + p~\Pi\indices{^i_j}\,, \label{eqn:emtdecompshear}
\eea
where $\Pi\indices{^i_j}= \tau\indices{^i_j} - \delta\indices{^i_j} \tau\indices{^k_k} / 3 $ is the anisotropic stress. Previously we have calculated the background energy density and pressure. The fluctuating part can be calculated by subtracting the average,
\begin{equation}
    \Delta\indices{^{\mu}_{\nu}} = \tau\indices{^{\mu}_{\nu}}-\langle \tau\indices{^{\mu}_{\nu}}\rangle \,.
\end{equation}
Using Eqs.~(\ref{eq:einstein_perts1}-\ref{eq:einstein_perts_4}) the fluctuating part can be related to the components of Eqs.~(\ref{eqn:emtdecomp}-\ref{eqn:emtdecompshear}). Numerically, however, these are much more challenging to calculate, as they cannot  easily be written in terms of the initial spectrum of fluctuations. We therefore take a phenomenological approach to the PGW perturbations, treating them as an effective Parameterized Post-Friedmann (PPF) fluid. 

The PPF framework is usually used in `smooth' dark energy models~\citep{Fang:2008sn}, but has several properties useful to model PGW perturbations. Firstly, it is able cross the $w=-1$ divide, which occurs for PGW oscillations after entering the horizon. Secondly, it is designed to conserve energy and momentum on large scales, where PGWs behave like positive curvature with $w_\trm{gw}=-1/3$. Finally, on small scales the PPF fluid is designed to be smooth compared to cold dark matter, which one would expect for PGWs due to the pressure support with $w_\trm{gw}=1/3$. In our approach we use the default PPF parameters in \textsc{camb}, and leave a more detailed study of PGW perturbations to future work.

One point worth mentioning is that in the SWA, PGWs are modelled as an effective neutrino species, with a hierarchy of moments describing the perturbations. In the second-order method no such hierarchy exists, and the fluid is described by its effective energy-momentum tensor. There are therefore two main observable differences expected compared to the shortwave treatment; (1) due to the background evolution, and (2) in the treatment of perturbations. 

\subsection{Observables}

\begin{figure}
\capstart
\includegraphics[width=\linewidth]{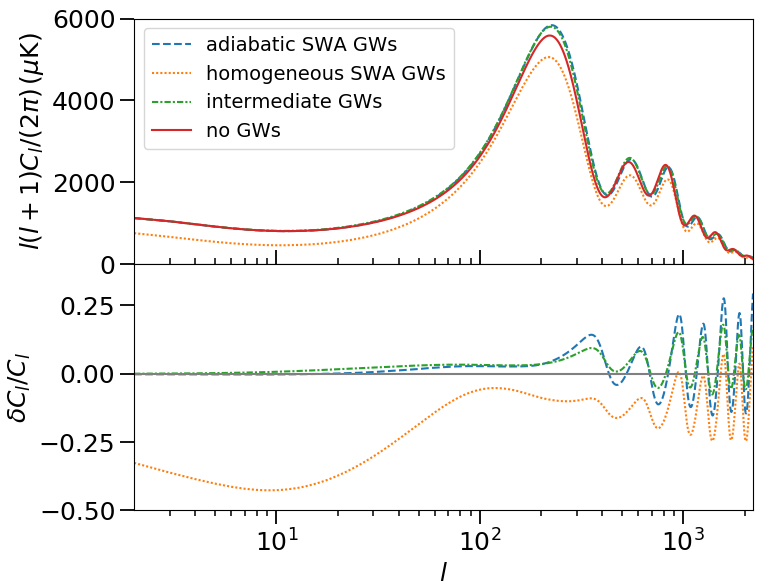}
\caption{Top panel: The CMB temperature power spectrum for $\Omgw h^2= 5.6\times10^{-6}$ with adiabatic shortwave approximation (SWA) initial conditions (blue dashed), homogeneous SWA initial conditions (orange dotted) (as detailed in section \ref{sec:high_frequencies}) and using the intermediate frequency method of section \ref{sec:non-shortwave} (green dot-dash). Bottom panel: the fractional difference in the CMB power spectrum due to gravitational waves as described above when compared to the case where there are no PGWs. }
\label{fig:cmb_power_spectrum_comp}
\end{figure}

The effects of high-frequency gravitational waves (as in section \ref{sec:high_frequencies}) on cosmological observables can be compared to the intermediate gravitational wave analysis of this section.  Figure \ref{fig:cmb_power_spectrum_comp} shows the CMB power spectrum for $\Omgw h^2=5.6\times10^{-6}$ for SWA gravitational waves with adiabatic and homogeneous initial conditions. The intermediate gravitational waves are shown for the same density with representative parameters of $n_t=3$ and $k_\text{min},\,k_\text{max}=0.1,\,1\, {\rm Mpc}^{-1}$ and no anisotropic stress. The intermediate analysis changes the temperature anisotropies in similar ways to the adiabatic SWA gravitational waves. 

\begin{figure}
\capstart
\includegraphics[width=\linewidth]{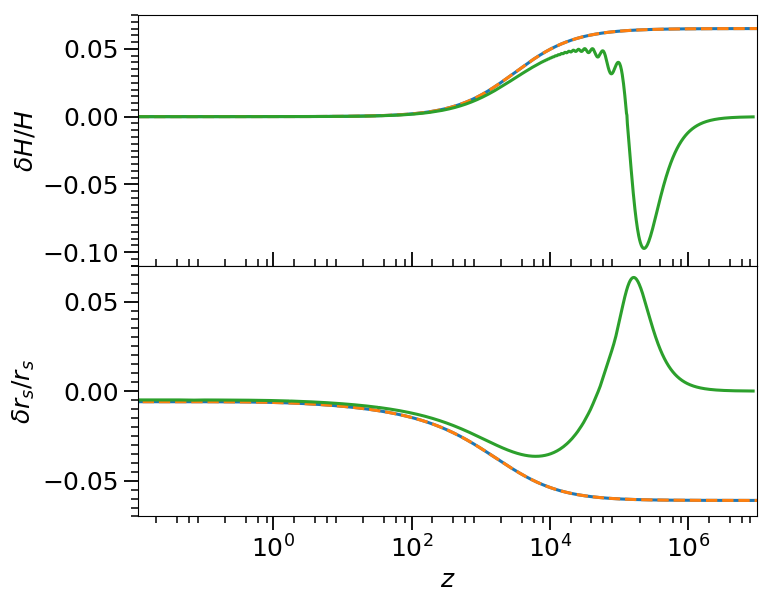}
\caption{Top panel: The fractional change in the Hubble rate as a function of redshift. The SWA analysis of section \ref{sec:high_frequencies} is used with adiabatic (blue) and homogeneous (orange dashed) initial conditions. The effects due to the intermediate frequency analysis of section \ref{sec:non-shortwave} is shown in green. When the equation of state of the intermediate frequency analysis goes negative a reduction in the Hubble rate occurs in contrast to the increase at early times seen for the shortwave analysis. Bottom panel: The fractional change in the size of the comoving sound horizon with line styles as above. }
\label{fig:hubble_rate_sound_horizon_comp}
\end{figure}

The fractional changes in the Hubble rate, $H(z)$, and the scale of the sound horizon, $r_s$, are shown in figure \ref{fig:hubble_rate_sound_horizon_comp}. As expected, due to these quantities only depending on the background and not the perturbations, the adiabatic and homogeneous high-frequency gravitational waves have identical affects on these parameters. The intermediate gravitational waves increase the Hubble rate similarly to the SWA result when dominated by high-frequency, $w_\trm{gw}=1/3$ modes but decreases the Hubble rate at high redshift when dominated by $w_\trm{gw}=-1/3$ modes. The same affect is seen in the scale of the sound horizon but with opposite sign and a small move to lower redshift. 

We note that the intermediate model shares some similarities with the axion-model that can potentially alleviate the Hubble tension~\citep{Poulin:2018cxd, Poulin:2018dzj}. In particular, there is an early dark energy (EDE) phase with $w_\trm{gw}=-1/3$ before a radiation phase with $w_\trm{gw}=1/3$. Even though the energy density is negative when $w_\trm{gw}=-1/3$, the sound horizon can still be reduced at the time of recombination. We leave the study of whether the PGW model can reduce the Hubble tension to future work.

This analysis assumes that the gravitational wave density is small enough that it can be calculated as a perturbation on a $\Lambda$CDM background. This was tested by iteratively recalculating the background including gravitational waves. Repeating this procedure until convergence shows an error of less than $0.01\%$ in  $H(z)$ over all $z$, for the maximum value of $\Omgw h^2$ allowed by data. We conclude that this is a small enough error to use the approximation that gravitational wave back-reaction can be calculated on a standard $\Lambda$CDM background. 

The observable consequences of primordial gravitational waves depends on the source function considered. Two source functions are considered here. So far a steep primordial power spectrum with tilt, $n_t\geq3$ has been used. This is motivated by the existing constraints and the possible sources in this region and is used for frequencies between $\sim 10^{-16}\,\text{Hz}$ and $\sim 10^{-15}\,\text{Hz}$. The second sources that will be considered are delta-function sources for specific frequencies. These give constraints that are independent of any assumptions about the spectrum of gravitational waves. These sources therefore give the upper-limit dependent only on the data and can be used as a consistency check on the steep sources as well as functioning as an independent constraint. 

\subsection{Parameter constraints}

To obtain limits on $\Omgw h^2$, a modified version of \textsc{camb} was integrated into \textsc{cobaya} to perform an MCMC analysis.  We use the same data as in section~\ref{sec:high_frequencies}, using an otherwise standard $\Lambda$CDM model. For PGWs, we choose $k_{\rm min}=0.1\,{\rm Mpc}^{-1}$, as below this the low-frequency constraint dominates, and $k_{\rm max}=1\,{\rm Mpc}^{-1}$, as above this the SWA can be used. We marginalise over the tensor spectral index, $n_{\rm t}$, in the prior range 3 to 5, where the lower limit is chosen to be compatible with the low-frequency constraint. The upper limit is chosen so as to include a range of short lasting early universe phenomena. Any production mechanism that produces gravitational waves in a short time frame will correspond to a large tilt. As examples, both cosmic strings and first order phase transitions can produce gravitational waves in the low to intermediate frequency regime \cite{Allen:1996vm,Binetruy:2012ze,Caprini:2015tfa,CapriniBmodeconversion}.

We obtain the following $95\%$ upper limits on the gravitational wave density parameter without neutrino anisotropic stress; 
\begin{equation}
    \Omgw h^2 < 8.4 \times10^{-7} \quad \text{(Second-order, no anisotropic stress)} \,.
\end{equation}
When including neutrino anisotropic stress the constraint has almost the same magnitude,
\begin{equation}
    \Omgw h^2 < 8.6 \times10^{-7}  \quad \text{(Second-order, with anisotropic stress)} \,.
\end{equation}
These are similar in magnitude to the shortwave adiabatic result and are tighter than the $B$-mode constraint for most of the region where the constraints overlap. These are integrated constraints and the constraint when neutrino anisotropic stress is included is shown as a horizontal line in figure \ref{fig:density} for the frequency range considered.

The values of the constraint on the gravitational wave density parameter for different wavenumbers in the range $0.02-0.5\,\text{Mpc}^{-1}$, when using delta-function sources and including neutrino anisotropic stress, are shown in table \ref{table:low:deltafunctionconstraints}. They are also plotted in figure \ref{fig:density} as stars. The constraint is extended to lower frequencies than the constraint for a steep source and weakens slightly as the frequency decreases but is of nearly the same magnitude for the region of overlap. 

\begin{table}
\centering
\def\arraystretch{1.9} 
\setlength{\tabcolsep}{17pt} 
 \resizebox{\textwidth}{!}{
\begin{tabular}{ c | c c c c c }
\hline\hline
    $k/\left(\text{Mpc}^{-1}\right)$ & $0.02$ & $0.045$ & $0.1$  & $0.18$ & $0.5$ \\
    $10^{17}\times f/\text{Hz}$ & $3.1$ & $7.0$ & $16$  & $28$ & $78$ \\
    $10^{7}\times \Omgw h^2$ ($95\%$ upper limit)  & $4.3$ & $7.7$ & $4.9$ & $6.1$ & $9.3$  \\ [1ex]
    \hline\hline
\end{tabular}
 }
\caption[Constraints on gravitational wave density for delta-function sources]{Constraints on the gravitational wave density for delta-function sources with wavenumber, $k$. The corresponding value of the frequency, $f$ is also shown. The constraint weakens slightly for lower frequencies but is of a similar magnitude to the adiabatic shortwave constraint for all $k$ considered.}
\label{table:low:deltafunctionconstraints}
\end{table}

\section{Conclusions} \label{sec:conclusions}

In this paper we have presented constraints on primordial gravitational waves from the CMB for the entire range of observable frequencies. This includes updated constraints from $B$-mode polarisation at the lowest frequencies, the shortwave approximation at high frequencies, and a new intermediate constraint that bridges the region of applicability of the two. These constraints are compatible at their extremities and provide the tightest current constraints in particular frequency ranges.

The constraint from low $\ell$ polarisation shows that peak sensitivity occurs for scales close to the horizon size at recombination, corresponding to $f\sim10^{-17}\, \text{Hz}$, with a gravitational wave density $\Omgw h^2 \sim 10^{-16}$. These limits become much weaker for $f \gtrsim 10^{-16}\, \text{Hz}$, and at $f \sim 3 \times 10^{-16}\, \text{Hz}$ a stronger result comes from the second-order back-reaction of gravitational waves. This allows us to place a limit of $\Omgw h^2 < 8.4\times10^{-7}$ in the absence of neutrino anisotropic stress and $\Omgw h^2 < 8.6\times10^{-7}$ when including neutrino anisotropic stress (both at $95\%$ confidence), in a previously unconstrained frequency region of $ 10^{-15}\, \text{Hz} \gtrsim f \gtrsim 3 \times 10^{-16}\, \text{Hz}$. At higher frequencies, $f \gtrsim 10^{-15}\, \text{Hz}$, we use the shortwave approximation (SWA) to update previous constraints and quantify the validity of the SWA using the intermediate approach, finding $\Omgw h^2 < 1.7\times10^{-6}$ for adiabatic initial conditions and $\Omgw h^2 < 2.9\times10^{-7}$ for homogeneous initial conditions (both at $95\%$ confidence). 

These constraints will be tightened by future ground and space based CMB observations from CMB-S4 and from polarisation via. LiteBIRD, CORE and PIXIE among others \cite{Abazajian:2016cmbs4,Hazumi:2019litebird,Delabrouille:2017core,Kogut:2011pixie}. These will result in an order of magnitude improvement in the measurement of extra relativistic degrees of freedom and an even greater improvement in the tensor-to-scalar ratio. Combining these with other cosmological observables promises to further illuminate the early Universe. 

There are several possibilities for future work. Due to the numerical challenges of calculating the fluctuations due to the second-order back-reaction, in this analysis we have treated them as an effective PPF fluid. In future work we plan to extend the line-of-sight CMB formalism to calculate these. It is worth noting though that, even in the shortwave limit, differences are expected compared to modelling them as an effective neutrino species with a hierarchy of moments. One further avenue might be investigating the possibility of PGWs alleviating the Hubble tension. The second-order model, with an appropriate source of PGWs, increases the relativistic degrees of freedom at recombination, thereby reducing the sound horizon, and having an early dark energy phase with $w_\trm{gw}=-1/3$. This would, however, require a non-standard source with a steep $n_\trm{t} \gtrsim 3$ spectrum peaking at $f \sim 10^{-15}\, \text{Hz}$.

\section*{Acknowledgements}
TJC would like to thank Finlay Noble Chamings, Karim Malik, Robert Brandenberger, Kouji Nakamura, Jens Chluba and Luke Hart for useful discussions. TJC is supported by a United Kingdom Science and Technology Facilities Council (STFC) studentship, AM is supported by a Royal Society University Research Fellowship and EJC is supported by STFC Consolidated Grant No. ST/P000703/1.

\appendix
\section{\label{sec:appendix}Solving the super-horizon gravitational wave equation of motion in a matter and cosmological constant background}

In this appendix conformal time is not used and dots denote cosmological time derivatives.

In a matter and cosmological constant background the scale factor,
\begin{equation}
    a(t)=\left(\frac{\Omega_\text{m}}{\Omega_\Lambda}\right)^{1/3}\left[\sinh\left(\frac{3}{2}H_0\sqrt{\Omega_\Lambda}t\right)\right]^{2/3} \,.
\end{equation}
So the gravitational wave equation of motion (compare to eq. (\ref{eq:eom})) in the absence of neutrino anisotropic stress,
\begin{equation}
    \ddot{D}(k,t)+3H(t)\dot{D}(k,t)+\frac{k^2}{a^2(t)}D(k,t)=0 \,,
\end{equation}
becomes,
\begin{equation}
   D''(\kappa,x)+2\coth x \,D'(\kappa,x)+\kappa^2(\sinh x)^{-4/3} D(\kappa,x)=0  \,,
\end{equation}
where,
\begin{equation}
    x=\frac{3}{2}H_0\sqrt{\Omega_\Lambda}t \,\,\,\,,\,\,\,\, \kappa= \left(\frac{\Omega_\Lambda}{\Omega_\text{m}}\right)^{1/3}\frac{2k}{3H_0\sqrt{\Omega_\Lambda}} \,,
\end{equation}
and primes denote differentiation with respect to $x$. $\kappa$ and $x$ are reduced wavenumber and time variables respectively. 

We solve the equation of motion in a power series for $\kappa^2$;
\begin{equation}
    D(\kappa,x)=D_0(x)+\kappa^2D_1(x) \,,
\end{equation}
because we are considering modes that are super-horizon at current (and near future) times. 

\subsection*{Background solution}
$D_0$ is the solution of the simpler equation,
\begin{equation}
   D_0''(x)+2\coth x \,D_0'(x)=0 \,,
\end{equation}
The general solution is $D_0(x)=\bar{D}_0-\alpha \coth x$. Imposing that the gravitational wave amplitude is finite as $x\to0$,
\begin{equation}
    D_0(x)=\bar{D}_0  \,.
\end{equation}
This constant is going to be set to $1$ in most cases. 

\subsection*{Perturbed solutions}
The equation of motion to order $\kappa^2$ is,
\begin{equation}
    D_1''(x)+2\coth x \,D_1'(x)+(\sinh x)^{-4/3} \bar{D}_0=0 \,.
\end{equation}

This can be solved in three separate regimes, low-$x$, intermediate-$x$ and high-$x$ and these solutions can be matched together at $x_a$ and $x_b$.

\subsubsection*{Low-x solution}
For small $x$ the equation of motion is,
\begin{equation}
    D_1''(x)+\frac{2}{x} \,D_1'(x)+\frac{\bar{D}_0}{x^{4/3}}=0 \,,
\end{equation}
with solution,
\begin{equation}
    D_1(x)=-\frac{9}{10}\bar{D}_0 x^{2/3} \,,
\end{equation}
where the initial condition is $D_1(0)=0$. 

\subsubsection*{Intermediate solution}

The intermediate solution is the most complicated and consequently we define new variables to simplify the solution. 

Expanding about the midpoint of the intermediate region, 
\begin{equation}
    \lambda=\frac{x_a+x_b}{2} \,,
\end{equation}
the intermediate solution is valid for more of the intermediate region than if either $x_a$ or $x_b$ was used. This results in the equation of motion becoming,
\begin{equation}
    D_1''(x)+2(\alpha+\beta x)D_1'(x)+\bar{D}_0(\gamma+\eta x)=0 \,,
\end{equation}
where,
\begin{align}
    \alpha&=\coth{\lambda}-\lambda\beta \,, &\beta=1-\coth^2{\lambda} \,, \nonumber \\
    \gamma&=\frac{3\sinh{\lambda}+4\lambda\cosh{\lambda}}{3(\sinh{\lambda})^{7/3}} \,, &\eta=-\frac{4\cosh{\lambda}}{3(\sinh{\lambda})^{7/3}} \,. 
\end{align}
Making the further definition,
\begin{equation}
    \bar{x}=\frac{\alpha+\beta x}{\sqrt{\beta}} \,,
\end{equation}
the intermediate solution for $D_1(x)$ is,
\begin{align}
    D_1(x)=\bar{D}_0\Bigg\{& C_1+\frac{C_2\sqrt{\pi} e^{\alpha^2/\beta}\erf{\bar{x}}-\sqrt{\beta}\eta x }{2\beta^{3/2}} \nonumber \\
    &+ \frac{(\beta\gamma-\alpha\eta)}{4\beta^3}\left[2\beta\bar{x}^2\, {}_{1}F_2\left(\{1,1\};\{3/2,2\};\bar{x}^2\right) -\pi\beta\erf{\bar{x}}\erfi{\bar{x}}\right] \Bigg\}
\end{align}
where $\erf{x}$ is the error function, $\erfi{x}$ is the imaginary error function, $_{p}F_q$ is the generalised hypergeometric function and the matching onto the low-$x$ solution at $x_a$ determines the coefficients $C_1$ and $C_2$. 

\subsubsection*{High-x solution}
For large $x$ the equation of motion becomes,
\begin{equation}
    D_1''(x)+2D_1'(x)+2^{4/3}\bar{D}_0 e^{-4x/3}=0 \,.
\end{equation}
The high$-x$ solution is,
\begin{equation}
    D_1(x)=C_3-\frac{C_4}{2}e^{-2x}+\frac{9\bar{D}_0}{2^{5/3}}e^{-4x/3} \,.
\end{equation}
$C_3$ and $C_4$ are determined by matching onto the intermediate solution at $x_b$. 

\subsection*{GW equation of state parameter}

\begin{figure}
\capstart
\includegraphics[width=\linewidth]{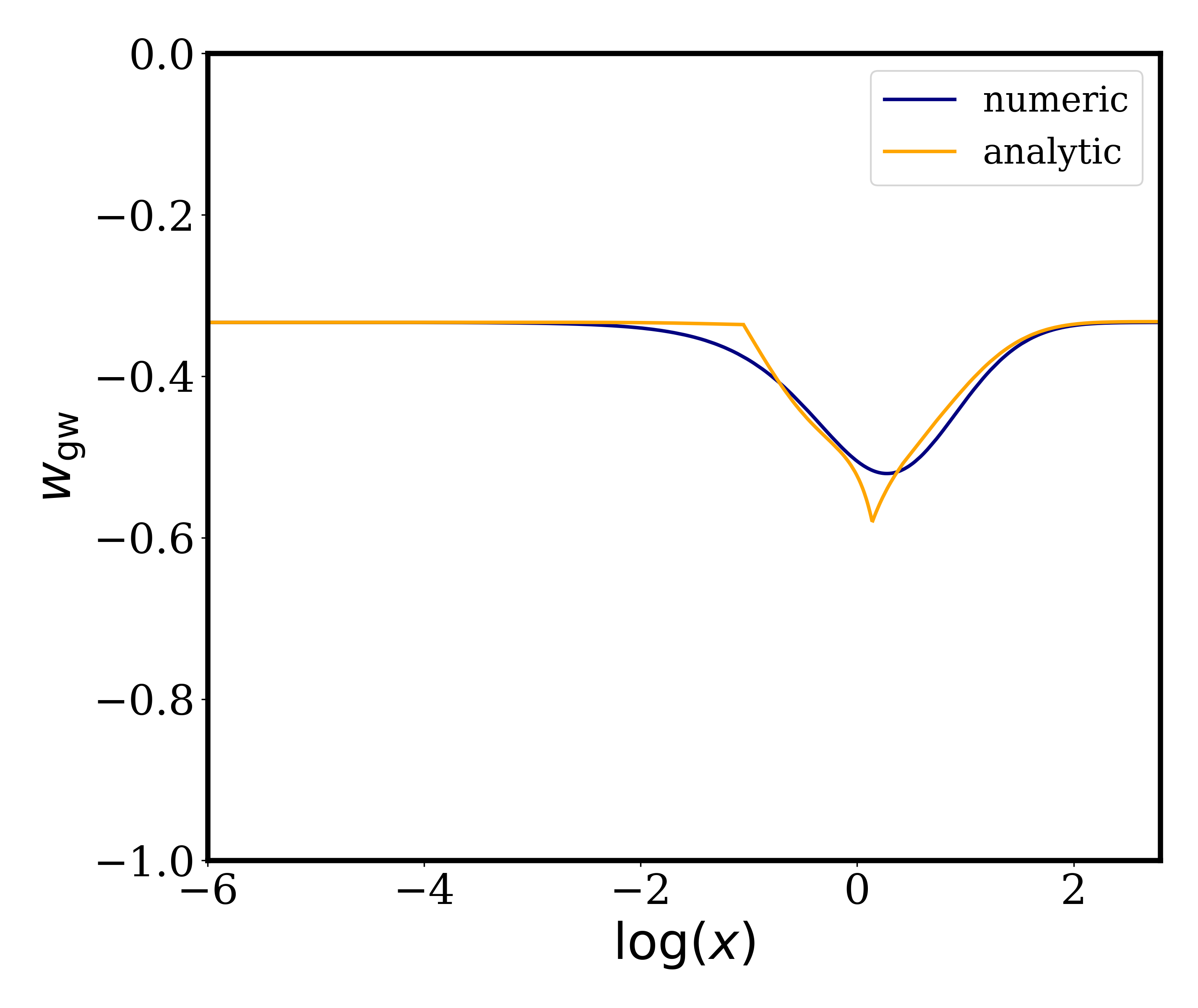}
\caption{The equation of state for gravitational waves as a function of $x$ for $\kappa=0.045$ (corresponding to $k=10^{-5}\, \text{Mpc}^{-1}$) from a numerical solution of the gravitational wave equation of motion (blue) and from an analytic solution found by matching solutions for small, intermediate and large $x$ (orange). The analytic solutions are matched together at $x_a=0.35$ and $x_b=1.15$ and verify the behaviour observed in the numerical solution. }
\label{fig:large_scale_eos}
\end{figure}

The gravitational wave equation of state parameter for a matter + cosmological constant background from the above analytics and from a numerical computation can be seen in figure \ref{fig:large_scale_eos}. The analytic solutions were matched at $x_a=0.35$ and $x_b=1.15$ to get the best agreement with the numerics. They confirm the fact that the equation of state departs from $-1/3$ for super-horizon GWs during the matter-cosmological constant transition but returns back to $-1/3$ when the cosmological constant comes to dominate. There are discontinuities due to imperfect matching of the solutions. Effectively the solutions are not of high enough order to fully encompass the behaviour in their specific regimes. This could be improved by using the intermediate solution twice and having four separate matched regimes but the analysis given here is sufficient to verify the numerical behaviour.

\bibliographystyle{JHEP.bst}
\bibliography{cmb_pgw}

\end{document}